\documentclass[lettersize,journal]{IEEEtran}
\usepackage{cite}
\usepackage{amsmath,amssymb,amsfonts}
\usepackage{algorithm}
\usepackage{algorithmic}
\usepackage{graphicx}
\usepackage{textcomp}
\usepackage{booktabs}
\usepackage{multirow}
\usepackage{xcolor}
\usepackage{url}
\usepackage{subcaption}
\def\BibTeX{{\rm B\kern-.05em{\sc i\kern-.025em b}\kern-.08em
    T\kern-.1667em\lower.7ex\hbox{E}\kern-.125emX}}
\begin{document}

\title{FlexP-SFT: A Flexible Aggregation-Free Framework for On-Device Personalized Split Federated Fine-Tuning of LLMs
}

\author{Jiaxiang Geng, Tianjun Yuan, Pengchao Han, Ying Gao, Xianhao Chen, and Bing Luo.
\vspace{-8mm}   
 \thanks{
   J. Geng, T. Yuan and B. Luo are with Duke Kunshan University, Suzhou, China (e-mail: \{jg645, ty117, bl291\}@duke.edu).
        
    P. Han is with Guangdong University of Technology, Guangzhou, China (e-mail: hanpengchao@gdut.edu.cn).

    Y. Gao is with Jade Bird Fire (JBF), China (e-mail: gaoying@jbufa.com).
    
    Xianhao Chen is with The University of Hong Kong, Hong Kong, China (e-mail: xchen@eee.hku.hk).
}
}

\maketitle

\begin{abstract}
To fine-tune large language models (LLMs) over private data, federated learning (FL) has emerged as a promising paradigm. However, the prohibitive memory and communication demands of LLMs render standard FL impractical for resource-constrained edge devices. While split federated learning (SFL) alleviates the computing burdens via model partitioning, existing frameworks still suffer from communication bottlenecks and straggler problem due to the parameter aggregation process. To address these challenges, we propose FlexP-SFT, a novel aggregation-free framework for personalized split federated fine-tuning, which fundamentally eliminates the client-side aggregation process. Crucially, to ensure robust training in the absence of global synchronization, we introduce a layer-flexible alignment strategy to balance personalization and generalization capabilities. We further formulate split-ratio selection as a resource-aware discrete optimization problem that jointly accounts for personalization accuracy and system cost. Our proposed scheme simultaneously enhances personalized performance, reduces communication overhead, and resolves the straggler problem. Extensive results show that FlexP-SFT substantially outperforms baselines in both accuracy and latency, and that the optimized split ratio achieves a better resource-accuracy trade-off than static or memory-only choices.
\end{abstract}

\vspace{-5mm} 
\section{Introduction}
LLMs, such as GPT \cite{radford2018improving, radford2019language, achiam2023gpt} and BERT \cite{devlin2018bert}, as well as more recent architectures \cite{warner2024smarter, bai2023qwen, team2024gemini, liu2024deepseek}, are large-scale machine learning models pre-trained on vast and diverse datasets \cite{bommasani2021opportunities}. These models are designed to capture broad and generalizable patterns across multiple domains, enabling strong performance on a wide range of tasks with minimal adaptation. To enhance their performance for specific users, fine-tuning is a widely used approach, especially when data distributions vary significantly across clients \cite{sun2019fine}. 
However, mainstream LLM fine-tuning approaches require uploading user data from personal devices to the centralized server, raising significant privacy concerns. Such centralized methods also fail to accommodate personalized applications, often resulting in suboptimal performance on user-specific tasks.

Federated learning (FL) enables multiple clients to collaboratively train a shared model while ensuring the privacy of their local datasets \cite{mcmahan2017communication, mammen2021federated}. This privacy-preserving property makes FL a promising paradigm for fine-tuning LLMs, addressing user privacy concerns while enhancing performance on downstream tasks \cite{zhuang2023foundation, ren2024advances}. However, traditional FL relies on the assumption that each client can locally host and train the entire model. This assumption becomes impractical for LLMs due to their prohibitive memory and computational demands. For instance, fully fine-tuning even a relatively compact 0.5-billion-parameter model necessitates approximately 10GB of peak memory to store model weights, gradients, and optimizer states, which is a requirement that vastly exceeds the capacity of typical consumer-grade edge devices.

To address the resource constraints, Split Federated Learning (SFL) \cite{thapa2022splitfed,10415235} has been proposed as a collaborative paradigm that partitions the model into client-side and server-side segments. A typical SFL workflow involves two distinct server roles: a Main Server for heavy computation and a Fed Server for global synchronization. Specifically, the client performs the forward pass up to a predefined cut layer and transmits the intermediate activations to the Main Server, which completes the remaining forward computation and executes back-propagation. Subsequently, the Fed Server aggregates the gradients or weights from the client-side models to update the global model. Generally, SFL architectures are categorized into SFL-V1 (separate server models) and SFL-V2 (shared server model). Crucially, due to the prohibitive memory cost on server of maintaining individual LLM copies for each client (SFL-V1), practical SFL deployments for LLM must adopt the shared server-side backbone strategy (SFL-V2), where a single server model serves all clients.

\begin{figure}[t]
\centering
\includegraphics[width=1\linewidth]{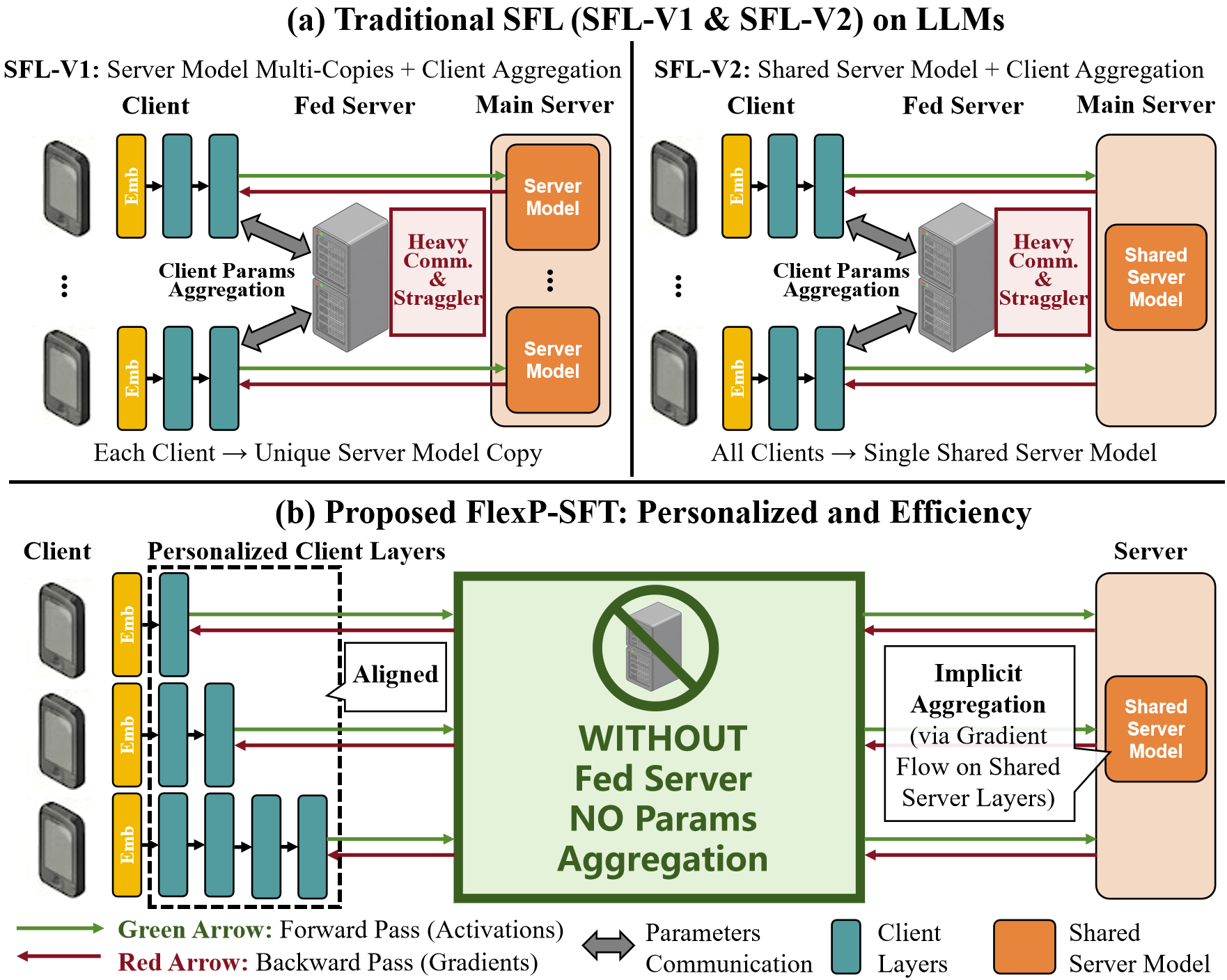}
\caption{Traditional SFL vs. the proposed FlexP-SFT.}
\label{pic:intro}
\vspace{-6mm}
\end{figure}

We refer to SFL-based LLM fine-tuning as Split Federated Fine-Tuning (SFT). Despite its potential, deploying existing SFT frameworks on heterogeneous edge devices is impeded by three critical bottlenecks \cite{fedbert,splitlora,sflllm,mesfl,11049940}: \textit{1) unresolved personalization-efficiency conflict:} Most current SFT approaches prioritize global convergence while overlooking personalized performance. Moreover, directly transplanting traditional personalized FL techniques (e.g., local adapters) introduces additional computational overheads, which are untenable for resource-constrained devices already saturated by LLM execution. \textit{2) prohibitive communication overhead:} Existing methods necessitate periodic aggregation of client-side models at the Fed Server. Although the LLM is split, the client-side segments still contain substantial parameters. Transmitting these high-dimensional updates over bandwidth-limited edge networks incurs massive latency, creating a significant bottleneck. \textit{3) exacerbated straggler problem:} Synchronous aggregation mandates that the server wait for all clients to complete uploads. In heterogeneous environments, high-performance devices face prolonged idle periods waiting for stragglers. The immense computational load of LLMs amplifies these processing disparities, leading to severe resource wastage and low system efficiency.

To summarize the above challenges and limitations, we identify the following key question:

\textit{How can we enable resource-constrained, heterogeneous edge devices to achieve high-performance personalized LLM fine-tuning, while eliminating the prohibitive communication overhead and straggler problem?}

To answer this key question, we take a bold step by fundamentally reimagining the SFL/SFT architecture: we eliminate the Fed Server and completely remove the client-side model aggregation process. Conventionally, aggregation is deemed indispensable in FL to mitigate local gradient bias and ensure global convergence. However, our empirical observations reveal a key insight specific to LLM fine-tuning: unlike training from scratch, pre-trained LLMs already possess robust general knowledge. Consequently, strict parameter synchronization for local layers is not only unnecessary but may dilute the model's ability to adapt to user-specific data.

By removing the aggregation procedure, our aggregation-free design achieves a \textit{"three birds with one stone"} effect, simultaneously resolving the three bottlenecks identified above: \textit{1) low-cost enhanced personalization}: Without the interference of global averaging, client-side models naturally evolve into highly specialized, personalized modules that better align with local distributions. \textit{2) zero communication for parameters aggregation:} By severing the transmission link between clients and the Fed Server, we eliminate the massive communication overhead associated with uploading high-dimensional client-side parameters. \textit{3) straggler-free asynchronous training:} The removal of the synchronization barrier transforms the system into a fully asynchronous workflow. Fast clients are no longer blocked by stragglers, maximizing system utilization and significantly accelerating the overall fine-tuning efficiency.

However, completely discarding global aggregation introduces a risk: without global constraints, client-side models may suffer from uncontrolled divergence, leading to overfitting on local data and a loss of general reasoning capability. To address this, we design a novel alignment mechanism aimed at balancing personalization and generalization. Crucially, this mechanism is designed to be layer-flexible. It seamlessly adapts to heterogeneous clients deploying varying numbers of local layers (e.g., adapting to different memory budgets), ensuring that clients with deeper local networks can fully utilize their capacity without drifting away from the global semantic space.

To sum up, in this paper, we propose \textbf{FlexP-SFT}, a flexible aggregation-free framework for on-device personalized SFT of LLMs. To the best of our knowledge, \textit{this is the first work to explore an aggregation-free SFL paradigm for LLMs, demonstrating that high-quality personalization can be achieved efficiently with low communication or computation cost}. Our main contributions are summarized as follows:\footnote{FlexP-SFT targets LLM fine-tuning without aggregating model parameters. Due to the massive scale and complexity of LLMs (e.g., transformers), theoretical convergence analysis remains an open challenge. We leave this as important future work, with potential directions inspired by related studies \cite{han2024convergence, pmlr-v162-castiglia22a}.}

\begin{itemize} \item \textbf{Aggregation-Free SFL Architecture for Efficiency}: We propose an aggregation-free SFL architecture eliminating the dependency on a fed server. By removing the client-side aggregation process, our design transforms the training workflow into a fully asynchronous paradigm. This structural shift enables intrinsic personalization at zero additional computational cost, reduces communication overhead and eradicates the straggler effect.

\item \textbf{Layer-Flexible Alignment for Balancing Personalization and Generalization}: To address system heterogeneity and preventing local overfitting in an aggregation-free setting, we introduce a \textbf{layer-flexible alignment mechanism}. This design allows heterogeneous clients to dynamically adjust their local model depth based on available resources. Crucially, we propose a novel regularization loss that implicitly aligns these diverse local representations with the global model's semantic space. This mechanism effectively balances personalization and generalization, ensuring that clients benefit from global knowledge without explicit parameter synchronization.

\item \textbf{Resource-Aware Split-Ratio Optimization}: We formulate the selection of the client-side split ratio as a discrete multi-variable optimization problem over Transformer layers. The formulation jointly considers personalization accuracy and system cost. We then design a profile-guided greedy algorithm that selects a feasible split for each heterogeneous client.

\item \textbf{Extensive Real-World Evaluation}: We implement a full-scale prototype of FlexP-SFT on a cluster of heterogeneous resource-constrained devices and conduct extensive experiments using representative commercial LLMs (e.g., decoder-only and encoder-only architectures). Compared with existing SFL baselines, our scheme improves accuracy by up to 5\%, while saving latency by up to 94\%. \end{itemize}

\textbf{Roadmap.} The remainder of this paper is organized as follows. Section~II reviews related works and presents our preliminary empirical analysis that motivates the aggregation-free design. Section~III details the proposed FlexP-SFT framework, the alignment strategy designed to balance personalization and generalization, and the resource-aware split-ratio optimization. Section~IV describes the experimental implementation and setup. Section~V presents a comprehensive evaluation of our approach in real-world scenarios. Finally, Section~VI concludes the paper.

\section{Related Works and Preliminaries}

\subsection{Federated On-device LLM Fine-Tuning}

\begin{figure}
\centering
\includegraphics[width=0.85\linewidth]{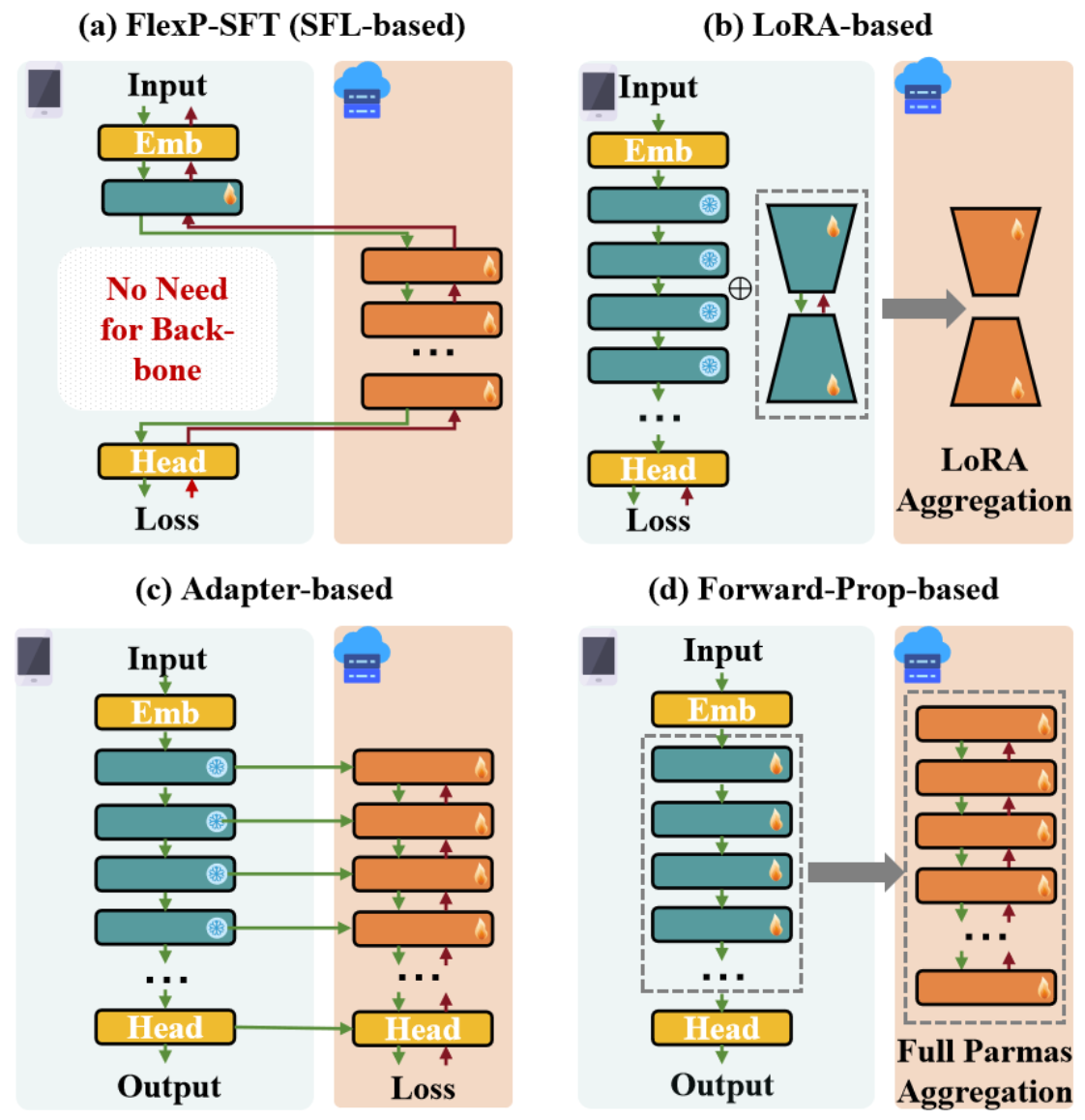}
\caption{Comparison among various approaches for federated on-device fine-tuning of LLMs.}
\label{pic:Comparison}
\vspace{-6mm}
\end{figure}

Researchers have explored various approaches for federated on-device fine-tuning of LLMs \cite{cho-etal-2024-heterogeneous, bai2024federated, wang2024flora, li2025mobillmenablingllmfinetuning, peng-etal-2024-pocketllm}, as shown in Fig.~\ref{pic:Comparison}. The methods except SFL-based methods suffer from significant limitations.
LoRA-based methods require each client device to locally update LoRA matrices and transmit them to the server for aggregation \cite{cho-etal-2024-heterogeneous, bai2024federated, wang2024flora}. Nevertheless, this approach still relies on synchronous updates across clients, making it vulnerable to straggler issues.
Adapter-based methods involve performing full forward propagation of the LLM on client devices and uploading the intermediate activations of each layer to the server \cite{li2025mobillmenablingllmfinetuning}. The server then trains adapters corresponding to each layer. However, this strategy incurs substantial communication overhead due to the transmission of large activation volumes. Moreover, computing adapter gradients on the server requires access to ground-truth labels, potentially leading to privacy leakage.
Forward-propagation-based methods attempt to reduce memory usage by estimating gradient directions solely via forward passes  \cite{peng-etal-2024-pocketllm}. While memory-efficient, such methods converge slowly and attempt to transmit the full parameters, thus significantly prolong wall-clock fine-tuning time.

Importantly, with the exception of SFL-based approaches, all aforementioned methods mandate that client devices load and execute the full LLM backbone. This requirement imposes a heavy computational burden that is often prohibitive for resource-constrained devices. In contrast, SFL-based frameworks are inherently more amenable to devices with limited memory and computational capabilities, as they allow clients to retain only a minimal fraction of the model parameters locally. For instance, in an extreme configuration, a client can retain solely the embedding layer while offloading all heavy hidden layers to the server.

\subsection{Existing SFL Methods on LLMs}

SFL has emerged as a promising paradigm for fine-tuning LLMs on resource-constrained edge devices. However, existing SFL methods on LLMs rely on client-side model aggregation to ensure global convergence. This dependency introduces significant system challenges, most notably substantial communication overhead for parameter transmission and the straggler problem caused by synchronous waiting.

For instance, FedBERT \cite{fedbert} splits the transformer architecture by retaining embedding and output layers on clients, yet it strictly requires aggregating these client-side parameters using FedAvg to maintain a global model. SplitLoRA \cite{splitlora} integrates Low-Rank Adaptation (LoRA) with SFL but explicitly defines a "Client-side LoRA Adapter Aggregation Stage," compelling clients to periodically upload their decomposition matrices to a local aggregation server for synchronization. SflLLM \cite{sflllm} necessitates a dedicated "Federated Server", which physically separate from the main computing server, specifically to aggregate trainable parameters after a set number of rounds. ME-SFL \cite{mesfl} relies on an edge server to periodically synchronize and aggregate full LoRA adapters from all users based on dataset weights.

Furthermore, existing SFL methods on LLMs typically prioritize training a single, generalized global model, often overlooking personalized performance. To facilitate direct aggregation, frameworks like SplitLoRA typically assume homogeneous client-side sub-models. Although recent works like ME-SFL attempt to address model heterogeneity (e.g., via varying LoRA ranks), they ultimately still aggregate local modules into a global knowledge base, failing to capture user-specific personalized performance.

\subsection{Personalized Federated Learning.} \ \
PFL \cite{arivazhagan2019federated, tan2022towards} framework consists of a parameter server and \( N \) local clients, where the \( n \)-th client holds its local dataset \( D_n \) with \( |D_n| \) samples.
Let \( w \in \mathbb{R}^k \) represent the global model parameters, and \( v_n \in \mathbb{R}^k \) denotes the personalized model parameters for client  \( n \)-th local model parameters for \( n \in [N] \), where \( [N] = \{1, 2, \dots, N\} \) represents the set of participating. The mathematical formulation of PFL can be expressed as:

\begin{equation}
\begin{aligned}
&\min_{v_n} L_n(v_n; w^*) = F_n(v_n) + \lambda \|v_n - w^*\|^2\\
&\text{subject to:} w^* = \arg \min_w \sum_{n=1}^{N} p_n L_n(v_n^*; w),
\end{aligned}
\end{equation}

The objective of PFL is to optimize the local objectives \( L_n(v_n; w^*) \) for each client, where \( w^* \) is the optimized global model. The local objective function consists of two key components: 1) the local empirical loss \( F_n(v_n) \) which measures the performance of the personalized model vn on the client's local dataset. 2) A regularization term \( \|v_n - w^*\|^2 \), which quantifies the divergence between the personalized model and the global model. \( \lambda \) and \( \{p_n\} \) are positive hyper-parameters in PFL. Specifically, \( \lambda \) controls the trade-off between personalization and global consistency. A larger \(\lambda \) enforces stronger alignment between local and global models. As \( \lambda \to \infty \), personalized FL reduces to standard SFL, where local models and the global model become identical. When \( \lambda = 0 \), personalized FL becomes to local learning, where each client trains its own model independently without considering the global model.

Representative methods include Ditto~\cite{li2021ditto}, which enforces personalization by regularizing local models towards global weights but necessitates iterative full-model transmission; APFL~\cite{deng2020adaptive}, which maintains dual sets of parameters to adaptively interpolate between local and global representations, thereby doubling memory usage; and FedRep~\cite{pmlr-v139-collins21a}, which keeps the classifier head local while aggregating the shared encoder, yet still obligates clients to perform computationally expensive full-model forward and backward passes.

However, while these strategies prove effective for traditional lightweight models, directly transplanting them to on-device LLMs imposes prohibitive computational and communication burdens. Edge devices are often already saturated by the heavy execution of LLMs. Consequently, imposing additional PFL overheads, such as storing dual large models as required by APFL or transmitting billion-parameter updates as in Ditto, becomes untenable for resource-constrained environments. Therefore, a lightweight, aggregation-free paradigm is essential to enable personalization without exacerbating the resource scarcity of edge devices.

\begin{table}[t]
\centering
\caption{Comparison between SFL (w/ Avg) and FlexP-SFT (w/o Avg) with Xavier and Pi4B}
\label{tab:transposed}
\vspace{1mm}
\resizebox{\linewidth}{!}{%
\begin{tabular}{ccc}
\toprule
\textbf{Metrics} & SFL (w/ Avg) & \textbf{FlexP-SFT (w/o Avg)} \\
\midrule
Final Accuracy  & 27.21\% & \textbf{28.65\%} \\
\begin{tabular}[c]{@{}c@{}}Overall Fine-tuning \\ Time  (SpeedUp)\end{tabular} & 5.89 hours (1.0×) & \textbf{1.31 hours (4.5×)} \\
\bottomrule
\end{tabular}%
}
\vspace{-4mm}
\end{table}

\begin{figure*}
\centering
\includegraphics[width=0.8\linewidth]{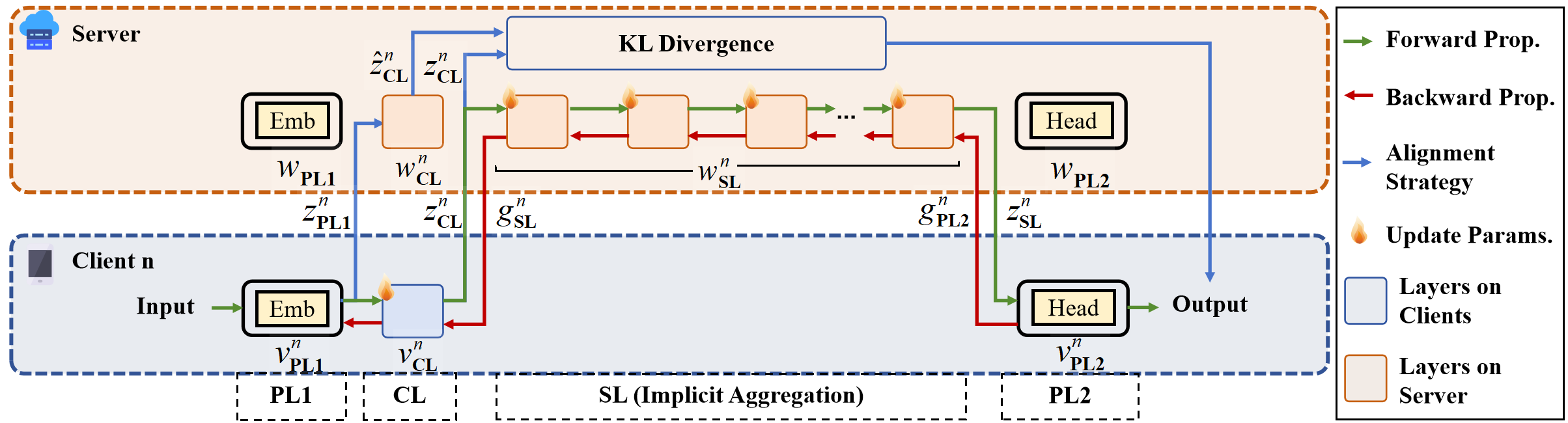}
\caption{The FlexP-SFT Framework (illustrated with one client $n$).}
\label{pic:FlexP-SFT-framework}
\end{figure*}

\subsection{Bold Attempt: Fine-Tuning LLMs w/o Parameters Aggregation.} \
A bold and novel attempt is to fine-tune LLMs \textbf{without} relying on a federated server or performing parameters aggregation (such as the FedAvg operation), as shown in Fig.~\ref{pic:intro}(b). This method removes the need for communication of client model parameters, replacing it with the exchange of activations and gradients only. Since the size of activations and gradients is significantly smaller than full model parameters, the communication burden is greatly reduced. In addition to reducing communication costs, the absence of FedAvg operations also means that clients are no longer required to synchronize their model updates. As a result, the straggler problem—which occurs when slower clients hold up faster clients during model aggregation—is effectively eliminated.

To evaluate the effectiveness of this approach, we conducted experiments on five heterogeneous devices, including Nvidia Xavier and Raspberry Pi 4B, using the BERT-Base model and the MMLU dataset. Clients were assigned heterogeneous datasets to reflect real-world scenarios. The experiments compared the performance of SFL (client-side model parameters aggregated with FedAvg) and our proposed method (aggregation-free). The results, summarized in Tab.~\ref{tab:transposed}, are remarkable:

\begin{itemize}
\item Our aggregation-free method achieves better personalized accuracy compared to SFL, demonstrating superior adaptability to client-specific data distributions. 

\item The reduced communication overhead and the elimination of the straggler problem also led to a 4.5x reduction in wall-clock time needed to fine-tune the model to the same level of accuracy. 
\end{itemize}

Based on these findings, we believe our approach is a promising direction for achieving better personalized performance in the fine-tuning of LLMs, while improving system efficiency and significantly shortening fine-tuning wall-clock time across heterogeneous client devices.

\section{Proposed Framework}
We present our \textbf{FlexP-SFT framework} as follows. We first introduce the FlexP-SFT framework overview in Sec.III-A, followed by the alignment strategy for balancing personalization and generalization in Sec.III-B and the resource-aware split-ratio optimization in Sec.III-C. We provide a detailed description of the forward and backward propagation workflow in Sec.III-D. Finally, we present two motivating scenarios in Sec.III-E.

\subsection{FlexP-SFT Framework Overview}

The FlexP-SFT framework involves a central server and $N$ clients, each with distinct computational capabilities and private datasets. The LLM is structured as a layered architecture, distributing processing responsibilities between clients and the server. \textbf{Figure~\ref{pic:FlexP-SFT-framework}} illustrates the interactions between these components. The complete model $w$ is divided into four parts: the Personalized Layers-1 (PL1), the Client Layers (CL), the Server Layers (SL) and the Personalized Layers-2 (PL2).

\noindent\textbf{Client-Side Model Components}: On the client side, each client stores and utilizes only a subset of the model’s layers, which consist of three distinct sets.
\begin{itemize}
\item \textbf{The Personalized Layers (PL1 and PL2)} are positioned at the beginning and end of the model, tailored to each client’s local data, and remain fully private. Both PL1 and PL2 are the minimum number of layers on clients to preserve the privacy of input raw data and labels, respectively. 

\begin{itemize}
\item \textbf{PL1} is responsible for initial feature extraction from raw inputs, includes components such as embedding layers. 
\item \textbf{PL2} handles the final transformations needed to produce the predictions, incorporating layers such as the output head layers. 
\end{itemize}

\item \textbf{Client Layers (CL)}: These layers lies between PL1 and PL2 and can be flexibly adjusted based on the client's resource constraints. Clients with higher computational capabilities can adopt deeper, more complex CL to extract richer local features, while resource-constrained clients can opt for shallower CL to reduce computational overhead. 
\end{itemize}

For client \(n\), the model parameters of PL1, CL and PL2 are represented as: $v_{\text{PL1}}$, $v_{\text{CL}}$ and $v_{\text{PL2}}$, respectively. The complete client model parameters for client \(n\) can thus be expressed as: 
\begin{equation}
v^n = \{v_{\text{PL1}}^n, v_{\text{CL}}^n, v_{\text{PL2}}^n\}.
\end{equation}

\noindent\textbf{Server-Side Model Components}: On the server side, one complete copy of the model is maintained, and only the \textbf{Server Layers (SL)} participate in gradient updates. The SL processes intermediate representations uploaded by clients, enabling the server to maintain a global view of the task without accessing any raw client data, thus preserving privacy. We denote the complete model on the server as $w$, and for client $n$, the shared SL is represented as $w^n_{\text{SL}}$. The server-side model components corresponding to the client-specific layers $v_{\text{PL1}}^n$, $v_{\text{PL2}}^n$, and $v_{\text{CL}}^n$ are denoted by $w_{\text{PL1}}$, $w_{\text{PL2}}$, and $w_{\text{CL}}^n$, respectively. Thus, the complete model on the server can be expressed as\footnote{Since each client has a different CL, their corresponding \( w_{\text{CL}}^n \) and \( w_{\text{SL}}^n \) on the server vary from the client's perspective. However, the general parameters on the server remain the same for all clients.}:  

\begin{equation}
\begin{aligned}
&w = \{w_{\text{PL1}}, w_{\text{CL}+\text{SL}}, 
w_{\text{PL2}} \}, \\
&\text{where} \ \ w_{\text{CL}+\text{SL}}=w_{\text{CL}}^n \cup w_{\text{SL}}^n, \quad \forall n \in N.
\end{aligned}
\end{equation}

For forward propagation of activations and backward propagation of gradients, the model structure used for client $n$ is exactly the same as $w$. This can be expressed as: 
\begin{equation}
\hat{v}^n = \{v_{\text{PL1}}^n, v_{\text{CL}}^n, 
w_{\text{SL}}^n, 
v_{\text{PL2}}^n \},
\end{equation}
which represents the complete model structure during the fine-tuning process for client $n$.

\begin{figure*}
\centering
\includegraphics[width=0.8\linewidth]{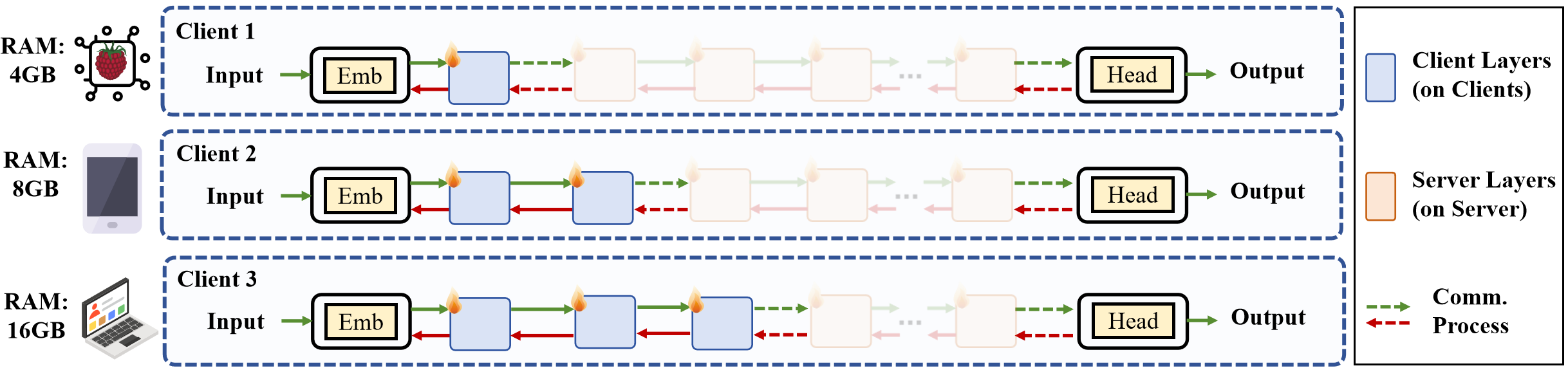}
\caption{Alignment strategy supports heterogeneous CL}
\label{pic:flexible}
\end{figure*}

\subsection{Layer-Flexible Alignment Strategy for Balancing Personalization and Generalization}

While the aggregation-free paradigm significantly enhances personalization and resolves both communication overhead and straggler issues, completely discarding global aggregation introduces a critical risk. Specifically, in the absence of global constraints, client-side models may suffer from uncontrolled divergence, leading to overfitting on local data and a loss of general reasoning capability. To address this personalization-generalization trade-off, FlexP-SFT integrates a layer-flexible alignment strategy that balances local adaptation with global model consistency. Inspired by Federated Knowledge Distillation (FedKD), we introduce a Kullback-Leibler (KL) divergence-based regularization to align the flexible layers between clients and the server \cite{ li2019fedmd, wu2022communication}. 
Specifically, the regularization term is defined as:

{
\begin{equation}
\mathcal{R}\left( v^{n}_{\text{CL}}, w_{\text{CL}}^n \right) = \text{KL}\left( P(z^{n}_{\text{CL}}) \,\|\, P(\hat{z}^{n}_{\text{CL}}) \right),
\label{regularization}
\end{equation}
}
where $\text{KL}(\bullet)$ represents the KL divergence function, \(P(\bullet)\) represents the probability distribution of the layers' output, and $\hat{z}^{n}_{\text{CL}}$ represents the output activations of $w_{\text{CL}}^n$ on server. 

To obtain \(\hat{z}^{n}_{\text{CL}}\), client \(n\) uploads \(z_{PL1}^n\) to the server, processes it through $w^n_{\text{CL}}$ to generate activations (see Fig.~\ref{pic:FlexP-SFT-framework}). The KL divergence is computed on the server, ensuring no additional computational burden on client devices. Although this approach introduces additional communication overhead (a few megabytes per round), it can be executed every few steps to reduce communication costs.

By minimizing the KL divergence, FlexP-SFT aligns the representations generated by the client’s CL parameters with those of the server, facilitating effective knowledge transfer and maintaining global model coherence. This prevents model drift that could result from excessive local personalization. Therefore, the local objective \(L_n\) for client \(n\) can be rewritten as:

\begin{equation}
L_n(v^n,w) = F_n(v^n) + \lambda R(v_{\text{CL}}^n, w_{\text{CL}}^n),
\label{loss_func}
\end{equation}
where \(\lambda\) controls the strength of the regularization, balancing the training loss and the regularization term. A larger $\lambda$ encourages the client model to stay closer to the global model, allowing the client to leverage knowledge aggregated from all devices. Conversely, a smaller $\lambda$ allows the client model to diverge more from the global model, enabling greater personalization to local data.
 \(F_n\) is the standard training loss (e.g., Mean Squared Error (MSE) or task-specific loss). \(R(v_{\text{CL}}^n, w_{\text{CL}}^n)\) is the KL-based regularization term from Eqn.~\eqref{regularization}.

Notably, our alignment strategy supports heterogeneous CL, allowing clients to deploy local sub-models of varying depths based on their specific device capabilities, as illustrated in Fig.~\ref{pic:flexible}. The remaining question is how each client should choose its CL depth under heterogeneous resource budgets. We formalize this split-ratio selection problem next.

\subsection{Resource-Aware Split-Ratio Optimization}

The layer-flexible alignment strategy enables FlexP-SFT to support heterogeneous CL, where different clients can retain different numbers of local Transformer layers according to their device capabilities. This flexibility raises a natural question: \textit{how should each client determine the number of CL layers to retain?} To formalize this decision, let $L$ denote the number of splittable Transformer layers after excluding PL1 and PL2. For client $n$, we define the normalized split-ratio decision variable as
\begin{equation}
Q_n \in \mathcal{Q}=\left\{0,\frac{1}{L},\frac{2}{L},\ldots,1\right\}.
\end{equation}
where $LQ_n$ represents the number of Transformer layers retained as CL on client $n$. The goal of this section is to determine $Q_n$ for heterogeneous clients in a resource-aware manner.

\noindent\textbf{Split-ratio trade-off.} The choice of \(Q_n\) determines how the splittable Transformer layers are distributed between client \(n\) and the server, thereby directly affecting both personalized performance and resource cost\footnote{In FlexP-SFT, the server maintains only one shared server-side model and usually has substantially larger computing resources than edge clients. Therefore, the resource cost considered in this section mainly focuses on client-side memory, computation latency, and energy consumption, which are the primary bottlenecks in heterogeneous on-device fine-tuning.}. Increasing \(Q_n\) allows client \(n\) to retain more trainable local layers, which strengthens its ability to adapt to the private data distribution. This trend is empirically validated in Table~\ref{tab:insight2}, where larger \(Q_n\) generally leads to higher final personalized accuracy after convergence. However, increasing \(Q_n\) also raises the client-side memory footprint, computation latency, and energy consumption, which may exceed the budgets of weak devices. Decreasing \(Q_n\), in contrast, reduces the local burden but leaves fewer trainable layers for personalization. Therefore, the split-ratio selection follows two key insights: \begin{itemize} \item \textbf{Insight 1: Larger \(Q_n\) improves personalization but increases local cost.} Retaining more CL layers gives the client stronger local adaptation capability, but the resulting memory, latency, and energy overhead must satisfy the device-side resource budgets. \item \textbf{Insight 2: Smaller \(Q_n\) reduces client-side cost but weakens personalization.} Retaining fewer CL layers makes FlexP-SFT more feasible for resource-constrained clients, but it also reduces the amount of client-side trainable capacity, which may limit personalized adaptation to local data. \end{itemize}

\noindent\textbf{Profiling table.}
Based on the above split-ratio trade-off, FlexP-SFT constructs a profiling table for each client \(n\) to decide the split ratio \(Q_n\). The profiling table contains two types of information: system resource cost and personalized-performance estimate. For the system resource cost, we focus on the client-side resource bottlenecks, including memory footprint and computation latency. Specifically, under split ratio \(Q_n\), we denote the memory footprint and computation latency of client \(n\) as \(M_n(Q_n)\) and \(T_n(Q_n)\), respectively. These two terms are normalized and combined into a unified resource-cost metric:
\begin{equation}
C_n(Q_n)
=
\omega_M \frac{M_n(Q_n)}{\overline{M}_n}
+
\omega_T \frac{T_n(Q_n)}{\overline{T}_n},
\end{equation}
where \(\overline{M}_n\) and \(\overline{T}_n\) denote the memory and latency budgets of client \(n\), while \(\omega_M\) and \(\omega_T\) control the relative importance of memory and latency.

For personalized performance, we use \(\widehat{A}_n(Q_n,\lambda)\) as a lightweight surrogate. This design is motivated by learning-curve extrapolation and multi-fidelity hyperparameter optimization, where early validation behavior is commonly used as a low-fidelity signal to estimate or rank the final performance of different training configurations~\cite{domhan2015speeding,falkner2018bohb,adriaensen2023efficient}. Specifically, for each candidate \(Q_n\), client \(n\) starts from the same initialized model and performs a short warm-up run for \(K_p\) local steps using the training objective in Eq.~\eqref{loss_func}. Let \(\mathcal{D}_n^{\mathrm{val}}\) denote a small local validation set, and let \(\mathcal{L}_{n,0}^{\mathrm{val}}\) and \(\mathcal{L}_{n,K_p}^{\mathrm{val}}(Q_n,\lambda)\) denote the validation losses before and after the warm-up run, respectively. We define the personalized-performance surrogate as the relative validation-loss reduction:
\begin{equation}
\widehat{A}_n(Q_n,\lambda)
=
\frac{
\mathcal{L}_{n,0}^{\mathrm{val}}
-
\mathcal{L}_{n,K_p}^{\mathrm{val}}(Q_n,\lambda)
}{
\mathcal{L}_{n,0}^{\mathrm{val}}+\epsilon
},
\end{equation}
where \(\epsilon\) is a small constant for numerical stability. A larger \(\widehat{A}_n(Q_n,\lambda)\) indicates that the corresponding split ratio brings faster local adaptation during the early fine-tuning stage, and is therefore expected to yield better personalized performance after convergence. This design avoids fully training every candidate split ratio before optimization. Therefore, each entry in the profiling table can be represented as
\begin{equation}
\mathcal{P}_n(Q_n)
=
\left\{
C_n(Q_n),\widehat{A}_n(Q_n,\lambda)
\right\},
\end{equation}
which provides the necessary cost-performance information for the subsequent split-ratio optimization.

\noindent\textbf{Optimization problem.}
Based on the profiling table, FlexP-SFT selects the split-ratio vector $\mathbf{Q}=\{Q_1,Q_2 \ldots,Q_N\}$
by jointly considering personalized performance and client-side resource cost. Since different clients have different data distributions and resource capacities, the personalized-performance estimate is client-specific rather than a fixed global target. Therefore, we formulate split-ratio selection as a weighted multi-objective optimization problem.

For client \(n\), the profiling table provides the resource-cost metric \(C_n(Q_n)\) and the personalized-performance surrogate \(\widehat{A}_n(Q_n,\lambda)\). We combine them into a utility function:
\begin{equation}
U_n(Q_n,\lambda)
=
\alpha \widehat{A}_n(Q_n,\lambda)
-
(1-\alpha) C_n(Q_n),
\end{equation}
where \(\alpha\in[0,1]\) controls the trade-off between personalized performance and resource efficiency. A larger \(\alpha\) gives higher priority to personalized accuracy, while a smaller \(\alpha\) favors lower client-side resource cost.

The overall split-ratio optimization problem is formulated as
\begin{equation}
\begin{aligned}
\max_{\mathbf{Q}} \quad &
\sum_{n=1}^{N} p_n U_n(Q_n,\lambda) \\
\mathrm{s.t.} \quad &
M_n(Q_n) \le \overline{M}_n,\quad \forall n\in\{1,\ldots,N\}, \\
&
T_n(Q_n) \le \overline{T}_n,\quad \forall n\in\{1,\ldots,N\}, \\
&
Q_n \in \mathcal{Q},\quad \forall n\in\{1,\ldots,N\}.
\end{aligned}
\tag{P1}
\end{equation}
Here, \(p_n=\frac{|\mathcal{D}_n|}{\sum_{m=1}^{N}|\mathcal{D}_m|}\) denotes the data-distribution weight of client \(n\), where \(|\mathcal{D}_n|\) is the number of local training samples on client \(n\). The objective maximizes the weighted utility across all clients, encouraging FlexP-SFT to assign larger split ratios to clients that can benefit more from additional local layers, while avoiding excessive memory and latency costs on resource-constrained devices. Since \(\mathcal{Q}\) is discrete due to Transformer-layer granularity, Problem~(P1) is a discrete multi-variable optimization problem.

\noindent\textbf{Solving P1.}
To obtain the optimal split-ratio set \(\mathbf{Q}^{\star}\), FlexP-SFT adopts a profile-guided enumeration method over the feasible candidates of each client. Although Problem~(P1) is discrete due to the layer-level candidate set \(\mathcal{Q}\), it does not require exhaustive search over all \(|\mathcal{Q}|^N\) split-ratio combinations. This is because the objective and the feasibility constraints are decomposable across clients. Therefore, FlexP-SFT solves P1 with a profile-guided selection procedure, as shown in Algorithm~\ref{alg:q_selection}. For each client, the algorithm first removes the split ratios that violate its memory or latency budget. It then computes the utility value of each remaining candidate based on the profiled personalized-performance surrogate and client-side resource cost. Finally, the candidate with the maximum utility is selected as the split ratio for that client. Since each client is processed independently, the algorithm only needs to scan the candidate set once for each client, resulting in a complexity of \(O(N|\mathcal{Q}|)\). The data-distribution weight \(p_n\) scales the contribution of client \(n\) in the global objective, but does not change the per-client maximizer because \(p_n>0\).

\begin{algorithm}[t]
\caption{Profile-Guided Split-Ratio Selection}
\label{alg:q_selection}
\begin{algorithmic}[1]
\REQUIRE Candidate set \(\mathcal{Q}\); profiling table \(\mathcal{P}_n(Q)\) with \(M_n(Q)\), \(T_n(Q)\), \(C_n(Q)\), and \(\widehat{A}_n(Q,\lambda)\); resource budgets \(\overline{M}_n\), \(\overline{T}_n\); trade-off weight \(\alpha\).
\ENSURE Optimal split-ratio set \(\mathbf{Q}^{\star}=\{Q_1^{\star},\ldots,Q_N^{\star}\}\).

\FOR{each client \(n\in\{1,\ldots,N\}\)}
    \STATE Initialize the feasible candidate set \(\mathcal{Q}_n=\emptyset\).
    
    \FOR{each candidate \(Q\in\mathcal{Q}\)}
        \IF{\(M_n(Q)\le \overline{M}_n\) and \(T_n(Q)\le \overline{T}_n\)}
            \STATE Add \(Q\) to \(\mathcal{Q}_n\).
        \ENDIF
    \ENDFOR
    
    \IF{\(\mathcal{Q}_n=\emptyset\)}
        \STATE \textbf{return} infeasible.
    \ENDIF
    
    \FOR{each candidate \(Q\in\mathcal{Q}_n\)}
        \STATE Compute the utility score:
        \[
        U_n(Q,\lambda)
        =
        \alpha \widehat{A}_n(Q,\lambda)
        -
        (1-\alpha)C_n(Q).
        \]
    \ENDFOR
    
    \STATE Select the best feasible split ratio:
    \[
    Q_n^{\star}
    =
    \arg\max_{Q\in\mathcal{Q}_n}
    U_n(Q,\lambda).
    \]
\ENDFOR

\STATE \textbf{return} \(\mathbf{Q}^{\star}=\{Q_1^{\star},\ldots,Q_N^{\star}\}\).
\end{algorithmic}
\end{algorithm}

\subsection{FlexP-SFT Workflow: Efficient Forward and Backward Propagation}

For client $n$, the fine-tuning workflow in FlexP-SFT can be formulated as follows:
{
\begin{equation}
\begin{aligned}
&\textbf{Forward propagation (activations $z$):} \\
&\quad \quad \quad \quad x_n \rightarrow v_{\text{PL1}}^n \xrightarrow{z^n_{\text{PL1}}} v_{\text{CL}}^n \xrightarrow{z^n_{\text{CL}}} w_{\text{SL}}^n \xrightarrow{z^n_{\text{SL}}} v_{\text{PL2}}^n \\
&\textbf{Backward propagation (gradients $g$):} \\
&\quad \quad \quad \quad v_{\text{PL1}}^n \xleftarrow{} v_{\text{CL}}^n \xleftarrow{g^n_{\text{SL}}} w_{\text{SL}}^n \xleftarrow{g^n_{\text{PL2}}} v_{\text{PL2}}^n \leftarrow L_n
\end{aligned}
\end{equation}}

\textbf{Forward Propagation}: Forward propagation in FlexP-SFT is executed in three stages—client $\to$ server $\to$ client, ensuring that the computationally intensive layers are processed on the server, while client-specific layers remain private on the client side.

\begin{itemize}

\item Client Processing: The client processes its local input \(x_n\) through \(v^{n}_{\text{PL1}}\) to obtain activations \(z^n_{\text{PL1}}\), followed by \(v^{n}_{\text{CL}}\) to generate activations \(z^n_{\text{CL}}\). These intermediate activations \(z^n_{\text{CL}}\) are then transmitted to the server.

\item Server Processing: The server receives \(z^n_{\text{CL}}\), processes them through \(w^{n}_{\text{SL}}\), and produces another set of activations \(z^n_{\text{SL}}\). These processed activations are then sent back to the client.

\item Final Client Processing: The client processes \(z^n_{\text{SL}}\) through \(v^{n}_{\text{PL2}}\) to generate the final predictions \(\hat{y}_n\).
\end{itemize}

\textbf{Backward Propagation}: The backward pass mirrors the forward propagation with a three-stage flow in the reverse direction—client $\to$ server $\to$ client—to ensure proper gradient computation across the entire model. This design enables all relevant layers, both shared and client-specific, to be updated effectively.

\begin{itemize}

\item Client Processing: The client computes its local loss \( L_n \) and initiates backward propagation through \( v^{n}_{\text{PL2}} \), generating gradients \( g^n_{\text{PL2}} \). These gradients are then sent to the server.  

\item Server Processing: The server receives \( g^n_{\text{PL2}} \), updates its parameters in \( w^{n}_{\text{SL}} \), and computes the necessary gradients \( g^n_{\text{SL}} \). These gradients are then transmitted back to the client.

\item Final Client Processing: The client updates \( v^{n}_{\text{CL}} \) by backward propagating \( g^n_{\text{SL}} \).  
\end{itemize}

The advantages of the proposed framework are as follows:

\begin{enumerate}
\item  \textbf{Overcoming Communication Overhead:} During the fine-tuning process, only \(z_{\text{CL}}^n\), \(z_{\text{SL}}^n\), \(g_{\text{PL2}}^n\), \(g_{\text{SL}}^n\) and \(z_{\text{PL1}}^n\) are transmitted between the client and server. These activations and gradients are significantly smaller in size compared to model parameters. 

\item \textbf{Mitigating the Straggler Problem:} On the server side, there is no parameter aggregation operation. The server-side procedure can run as a service without requiring synchronization of packages sent by clients. This asynchronous processing eliminates the need for clients to wait for each other, solving the straggler problem, reducing idle time, and accelerating the overall fine-tuning process.  

\item \textbf{Enhanced Personalized Performance:} Instead of direct model parameter aggregation, the proposed framework improves personalized performance by better adapting the model to individual client data. 
\end{enumerate}

\subsection{Motivating Scenarios}
In this section, we showcase how FlexP-SFT can be applied in the two motivating scenarios.

\subsubsection{Campus Health Assistant}
The rapid development of wearable devices has enabled students to continuously collect rich health-related data through smartwatches, including calorie consumption, step counts, heart rate, and sleep duration. In a campus context, such as physical education classes, these data can help instructors design training programs that better match students’ health conditions and fitness levels, ultimately improving training outcomes \cite{FedCampus}. We envision using these on-device data to fine-tune a health assistant LLM that generates personalized exercise and course recommendations for both students and teachers. However, these data often contain sensitive personal information, so that students may be unwilling to upload them to a central server for model training. To address this challenge, our proposed FlexP-SFT framework enables such on-device fine-tuning in a privacy-preserving and device-heterogeneity manner, allowing students with diverse resource capacities to participate in the finetuning process with flexible computational and communication overhead and receiving accurate personalized health guidance.

\subsubsection{Personal Asset Management Assistant}
Smartphones store a wealth of financial data, including users’ spending habits, transaction histories, and budgeting records. These data are highly valuable for developing a personal asset management assistant powered by LLMs, which can provide customized financial planning, savings advice, and expense optimization strategies. However, since such financial information is extremely sensitive, users are often unwilling or even legally restricted to upload it to a central server for model fine-tuning. FlexP-SFT provides a practical solution by enabling secure on-device fine-tuning of the LLM while preserving user privacy. Through flexible model partitioning, FlexP-SFT accommodates devices with different computational capacities, allowing each user to participate in collaborative training with flexible resource overhead. In this way, users can benefit from personalized and privacy-preserving financial guidance.

\begin{figure}
\centering
\includegraphics[width=0.9\linewidth]{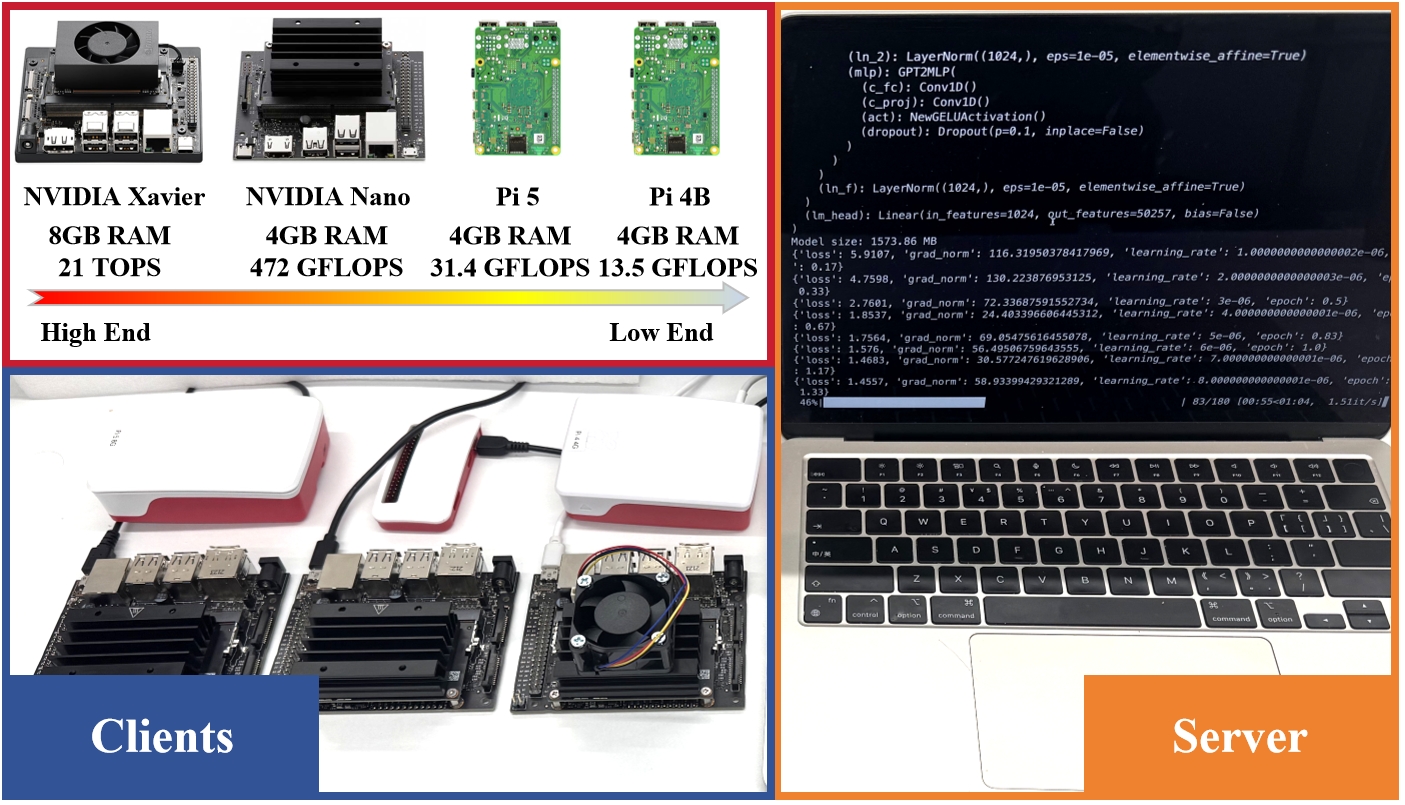}
\caption{Testbed of FlexP-SFT}
\label{pic:demo}
\end{figure}

\begin{table}[]
\centering
\caption{Client Devices List in Testbed}
\label{tab:device_all}
\resizebox{\linewidth}{!}{%
\begin{tabular}{cccccc}
\toprule
\begin{tabular}[c]{@{}c@{}}Clients\end{tabular} &
\begin{tabular}[c]{@{}c@{}}Device\end{tabular} &
\begin{tabular}[c]{@{}c@{}}Capacity\end{tabular} &
\begin{tabular}[c]{@{}c@{}}Memory\end{tabular} &
\begin{tabular}[c]{@{}c@{}}CPU\end{tabular} &
\begin{tabular}[c]{@{}c@{}}GPU\end{tabular} \\
\midrule
C1, C2 &
\begin{tabular}[c]{@{}c@{}}NVIDIA\\ Jetson Xavier\end{tabular} &
21 TOPS & 8 GB &
\begin{tabular}[c]{@{}c@{}}8-core\\ ARM CPU\end{tabular} &
\begin{tabular}[c]{@{}c@{}}1 Volta\\ GPU\end{tabular} \\

C3, C4 &
\begin{tabular}[c]{@{}c@{}}NVIDIA\\ Nano\end{tabular} &
472 GFLOPS & 4 GB &
\begin{tabular}[c]{@{}c@{}}1 quad-core\\ ARM CPU\end{tabular} &
\begin{tabular}[c]{@{}c@{}}1 Maxwell\\ GPU\end{tabular} \\

C5 &
\begin{tabular}[c]{@{}c@{}}Raspberry\\ Pi 5\end{tabular} &
31.4 GFLOPS & 16 GB &
\begin{tabular}[c]{@{}c@{}}1 quad-core\\ ARM CPU\end{tabular} & - \\

C6 &
\begin{tabular}[c]{@{}c@{}}Raspberry\\ Pi 4B\end{tabular} &
13.5 GFLOPS & 4 GB &
\begin{tabular}[c]{@{}c@{}}1 Cortex A72\\ ARM CPU\end{tabular} & - \\
\bottomrule
\end{tabular}%
}
\end{table}

\begin{table*}[]
\centering
\caption{Performance Compared with traditional FL methods 
(FA.: \underline{F}inal \underline{A}ccuracy, OFT.: \underline{O}verall \underline{F}ine-tuning \underline{T}ime)}
\label{tab:overall}

\begin{tabular}{clc cc cc cc}
\toprule
\multicolumn{3}{c}{Settings} & \multicolumn{2}{c}{\textbf{Ours}} & \multicolumn{2}{c}{FedAvg} & \multicolumn{2}{c}{SFL} \\
\cmidrule(lr){1-3} \cmidrule(lr){4-5} \cmidrule(lr){6-7} \cmidrule(lr){8-9}
Model & \#Para & Dataset & FA. (\%) & OFT. (h) & FA. (\%) & OFT. (h) & FA. (\%) & OFT. (h) \\
\midrule
BERT-Large     & 340M & \multirow{6}{*}{MMLU}     & \textbf{28.89} & \textbf{1.57}  & 27.55 & 128.33 & 27.38 & 27.33 \\
BERT-Base      & 110M &                           & \textbf{28.81} & \textbf{1.18}  & 27.53 & 31.63  & 27.38 & 5.70  \\
GPT2-Medium    & 355M &                           & \textbf{26.20} & \textbf{1.63}  & 25.45 & 105.12 & 25.59 & 31.87 \\
GPT2-Small     & 124M &                           & \textbf{25.18} & \textbf{1.27}  & 24.69 & 49.22  & 24.68 & 8.68  \\
Qwen3-0.6B  & 0.6B   &                           & \textbf{44.95}            & \textbf{1.54}           & 43.70        & 64.07       & 43.59      & 3.02       \\
Qwen3-1.7B  & 1.7B   &                           & \textbf{55.41}            & \textbf{3.09}           & 54.74        & 182.95      & 53.85      & 8.63       \\ 
\midrule
BERT-Large     & 340M & \multirow{6}{*}{MMLU-Pro} & \textbf{19.63} & \textbf{1.56}  & 16.85 & 129.78 & 19.11 & 28.65 \\
BERT-Base      & 110M &                           & \textbf{19.50} & \textbf{1.15}  & 15.75 & 42.74  & 19.06 & 4.31  \\
GPT2-Medium    & 355M &                           & \textbf{14.23} & \textbf{1.62}  & 13.98 & 108.45 & 13.87 & 30.18 \\
GPT2-Small     & 124M &                           & \textbf{11.69} & \textbf{1.14}  & 11.22 & 48.98  & 11.12 & 8.59  \\
Qwen3-0.6B     & 0.6B &                           & \textbf{25.84} & \textbf{1.52}  & 24.91 & 65.83  & 24.76 & 3.18  \\
Qwen3-1.7B     & 1.7B &                           & \textbf{35.72} & \textbf{3.12}  & 34.89 & 185.67 & 34.31 & 8.91  \\
\bottomrule
\end{tabular}
\end{table*}

\begin{table*}[]
\centering
\caption{Performance Compared with PFL methods 
(FA.: \underline{F}inal \underline{A}ccuracy, OFT.: \underline{O}verall \underline{F}ine-tuning \underline{T}ime)}
\label{tab:personalized_performance}

\resizebox{\linewidth}{!}{%
\begin{tabular}{clcclcccccccc}
\toprule
\multicolumn{4}{c}{Settings} & \multicolumn{2}{c}{\textbf{Ours}} & \multicolumn{2}{c}{APFL} & \multicolumn{2}{c}{Ditto} & \multicolumn{2}{c}{FedRep} \\
\cmidrule(lr){1-4} \cmidrule(lr){5-6} \cmidrule(lr){7-8} \cmidrule(lr){9-10} \cmidrule(lr){11-12}
Model & \#Para & Dataset & Pretrained Acc. (\%) & FA. (\%) & OFT. (h) & FA. (\%) & OFT. (h) & FA. (\%) & OFT. (h) & FA. (\%) & OFT. (h) \\
\midrule
BERT-Large     & 340M  & \multirow{6}{*}{MMLU} & 25.87 & \textbf{28.89} & \textbf{1.57} & 28.18 & 100.16 & 29.91 & 90.47  & 27.04 & 113.57 \\
BERT-Base      & 110M  &                        & 22.17 & \textbf{28.81} & \textbf{1.18} & 27.01 & 42.74  & 25.81 & 39.96  & 26.19 & 37.19  \\
GPT2-Medium    & 355M  &                        & 24.68 & \textbf{26.20} & \textbf{1.63} & 26.57 & 122.09 & 26.48 & 114.99 & 26.80 & 123.97 \\
GPT2-Small     & 124M  &                        & 22.32 & \textbf{25.18} & \textbf{1.27} & 25.04 & 45.53  & 24.74 & 43.32  & 24.80 & 41.59  \\
Qwen3-0.6B     & 0.6B  &                        & 42.10 & \textbf{44.95} & \textbf{1.54} & 44.41 & 72.68  & 44.67 & 69.35  & 44.20 & 75.12  \\
Qwen3-1.7B     & 1.7B  &                        & 52.60 & \textbf{55.41} & \textbf{3.09} & 55.02 & 207.18 & 55.18 & 198.74 & 54.91 & 214.62 \\
\bottomrule
\end{tabular}%
}
\end{table*}

\begin{table*}[]
\centering
\small
\caption{Performance Compared with On-device Fine-tuning Methods 
(Memory: Average memory consumed on each device, Comm.: Total communication throughput of each device)}
\label{tab:3rdbaseline}
\resizebox{0.75\textwidth}{!}{%
\begin{tabular}{ccccccc}
\toprule
\multicolumn{4}{c}{Settings} & \multicolumn{3}{c}{Metrics} \\
\cmidrule(r){1-4} \cmidrule(l){5-7}
Model & Dataset & Pretrained Acc. & Method & Accuracy (\%) & Memory (GB) & Comm. (GB) \\
\midrule
\multirow{4}{*}{ModernBERT-Base} 
  & \multirow{4}{*}{MMLU} 
  & \multirow{4}{*}{25.21} 
  & \textbf{Ours}      & \textbf{28.89} & \textbf{0.57} & \textbf{1.07} \\
  &                    &                & FLoRA         & 26.63 & 1.91 & 3.78 \\
  &                    &                & MobiLLM       & 27.88 & 1.26 & 5.88 \\
  &                    &                & PocketLLM     & 27.34 & 1.40 & 116.99 \\
\bottomrule
\end{tabular}%
}
\vspace{-4mm}
\end{table*}

\section{Experiment Setup}

\subsection{FlexP-SFT Testbed}

\subsubsection{Platform}

FlexP-SFT is implemented on a testbed as illustrated in Fig.~\ref{pic:demo}, consisting of a server and six clients. On the server side, an NVIDIA RTX 3090 with 24GB of memory capacity is utilized. On the client side, we consider four types of mobile devices with heterogeneous computational capacities, listed in Tab.\ref{tab:device_all}.
For communication between the server and clients, we use the WebSocket communication protocol and Wi-Fi 5 for wireless transmissions. Additionally, command-line tools such as nmcli and wondershaper are employed for monitoring network status and controlling wireless transmission rates.

\subsubsection{Implementation}Unlike typical fine-tuning methods for large models based on Transformers Tool Package, we developed the fine-tuning framework based on PyTorch. As detailed in Table~\ref{tab:device_all}, clients are numbered by device type for clarity: C1 and C2 are the two NVIDIA Jetson Xavier devices, C3 and C4 are the two NVIDIA Jetson Nano devices, C5 is the Raspberry Pi 5, and C6 is the Raspberry Pi 4B. This framework is designed to support the transmission of activations and gradients between the server and clients, forward and backward propagation on both the server and clients, and the heterogeneous deployment of adaptive layers across devices. The framework is compatible with any encoder/decoder-based LLM architecture and allows seamless integration of new devices.

\subsubsection{Datasets}We evaluated the performance of our proposed FlexP-FSL method on the Massive Multitask Language Understanding (MMLU) dataset \cite{hendrycks2020measuring} (4 options, across 57 domain tasks) and MMLU-Pro Dataset \cite{wang2024mmlu} (10 options, across 14 domain tasks). 
In our experiments, each client was assigned data corresponding to exactly one domain task. This setup represents an extremely challenging data distribution scenario, as it introduces both label non-i.i.d. and task non-i.i.d. characteristics. The evaluation results were obtained by averaging the outcomes across various distribution strategies.

\subsubsection{Models}The experiments were conducted using 7 language models for full fine-tuning: BERT-Large-Uncased (340M), BERT-Base-Uncased (110M), ModernBERT-Base (149M), GPT-Medium (355M), GPT-Small (124M), Qwen3-0.6B and Qwen3-1.7B. To ensure a comprehensive assessment, each model underwent a rigorous testing protocol spanning thousands of data batches to converge. For memory-intensive methods like FedAvg, devices with 4GB of memory are insufficient to support training. To address this limitation, we developed a mirroring tool that replicates the computational capabilities of client devices on the server for simulation. This allows for convenient performance comparison across different experimental setups.

\subsubsection{Hyper-parameters}
We denote the learning rate by $\eta$ to distinguish it from the accuracy--resource trade-off weight $\alpha$ in Problem~(P1). GPT-Medium and GPT-Small use Adam with $\eta=4\times10^{-9}$, while BERT-Large-Uncased and BERT-Base-Uncased use Adam with $\eta=5\times10^{-8}$. The alignment weight is initialized as $\lambda=0.1$. Table~\ref{tab:q_hyperparameters} lists the additional settings used by the resource-aware split-ratio selector.

For each client and candidate $Q_n$, we copy the same initialized checkpoint, reserve 10\% of the local training data as $\mathcal{D}_n^{\mathrm{val}}$, and run $K_p=50$ warm-up steps using the same optimizer, learning rate, and alignment loss as full fine-tuning. We record the validation loss immediately before warm-up, $\mathcal{L}_{n,0}^{\mathrm{val}}$, and after the 50th step, $\mathcal{L}_{n,50}^{\mathrm{val}}(Q_n,\lambda)$. Their relative reduction is then substituted directly into the definition of $\widehat{A}_n(Q_n,\lambda)$ in Sec.~III-C. Test labels and final converged accuracy are not used by Algorithm~\ref{alg:q_selection}. Unless otherwise stated, we set $\omega_M=\omega_T=0.5$ and $\alpha=0.7$; the sensitivity study varies $\alpha$ from 0.1 to 0.9.

\subsection{Algorithms Comparison}
In our experimental evaluation, we compare against 3 groups of baselines.

\subsubsection{Comparison with traditional FL methods}The first group includes two of the most widely adopted federated learning architectures: SFL \cite{thapa2022splitfed} (SFL-V2) and FedAvg \cite{mcmahan2017communication}. These methods represent the mainstream approaches in traditional federated learning and serve as strong general-purpose baselines.

\subsubsection{Comparison with PFL methods}The second group consists of state-of-the-art personalized federated learning methods, including Ditto \cite{li2021ditto}, APFL \cite{deng2020adaptive}, and FedRep \cite{pmlr-v139-collins21a}. These baselines are designed to improve personalization in federated settings and thus offer a meaningful point of comparison to evaluate the personalized performance of FlexP-SFT.

\subsubsection{Comparison with federated on-device fine-tuning
of LLM methods} The third comparison group includes recent methods developed for federated on-device fine-tuning of LLMs, including FLoRA \cite{wang2024flora} (LoRA-based), MobiLLM \cite{li2025mobillmenablingllmfinetuning} (Adapter-based), and PocketLLM \cite{peng-etal-2024-pocketllm} (Forward-Prop-based), which have been shown in Fig.~\ref{pic:Comparison}. We compare FlexP-SFT with these baselines to evaluate our efficiency and personalization performance on fine-tuning LLMs on devices.


\section{Results}

In this section, we present the key experimental results. In Sec. V-A, we first compare our method against the three groups of baseline methods introduced above to demonstrate its superiority. In Sec. V-B, we conduct ablation studies, including the effect of \(Q\), the proposed resource-aware split-ratio selector, and the alignment coefficient. Finally, in Sec. V-C, we assess the robustness of FlexP-SFT, including its performance under communication dropout caused by network fluctuations and its ability to personalize for individual clients.

\begin{figure}[t]
  \centering
  \begin{subfigure}[t]{0.49\linewidth}
    \centering
    \includegraphics[width=\linewidth]{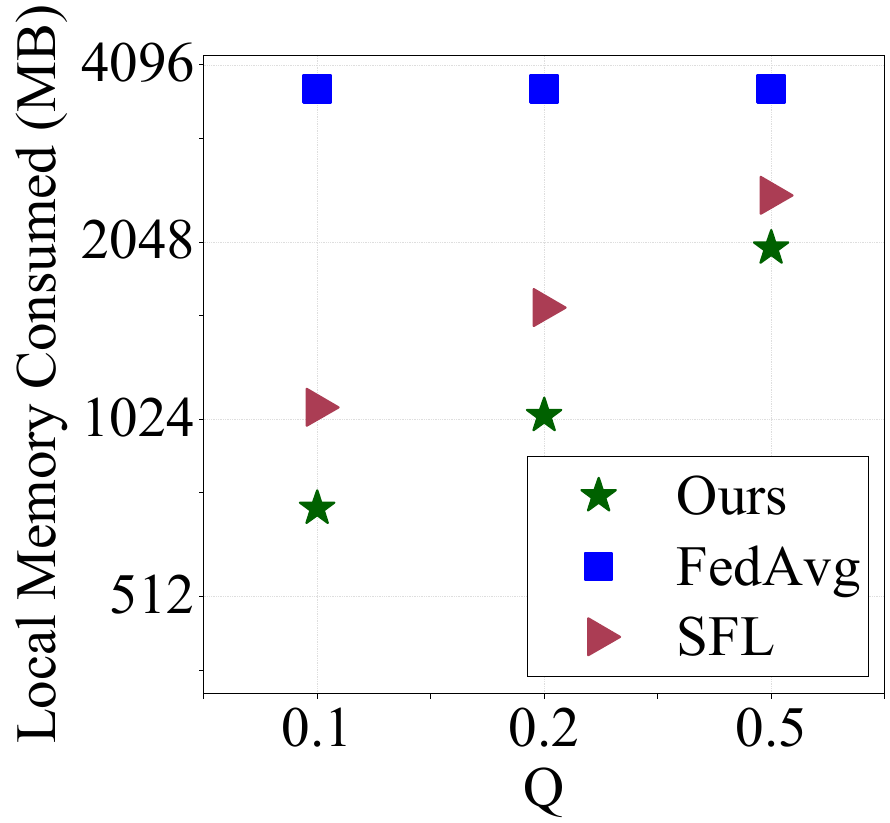}
    \caption{Local Memory}
    \label{fig:localmemory}
  \end{subfigure}
  \hfill
  \begin{subfigure}[t]{0.49\linewidth}
    \centering
    \includegraphics[width=\linewidth]{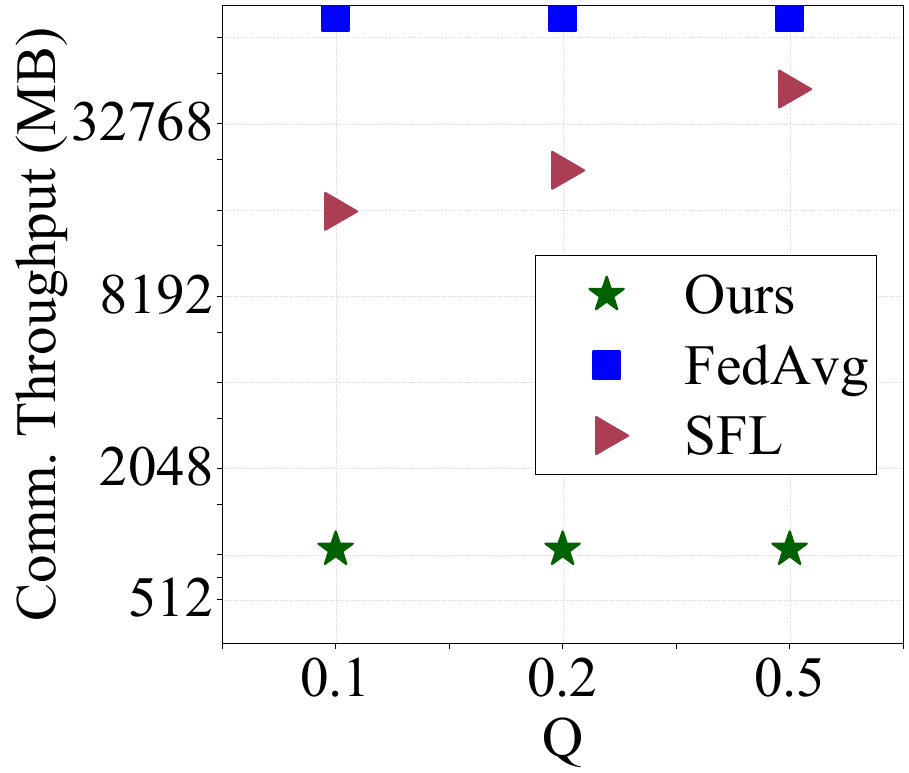}
    \caption{Throughput}
    \label{fig:commthroughput}
  \end{subfigure}
  \caption{Local memory and communication throughput.}
  \label{fig:memory_throughput}
\vspace{-4mm}
\end{figure}

\subsection{Comparison with Baselines}

\subsubsection{Comparison with traditional FL methods} 
\begin{figure}[t]
  \centering
    \begin{subfigure}[t]{0.47\linewidth}

      \centering
      \includegraphics[width=\linewidth]{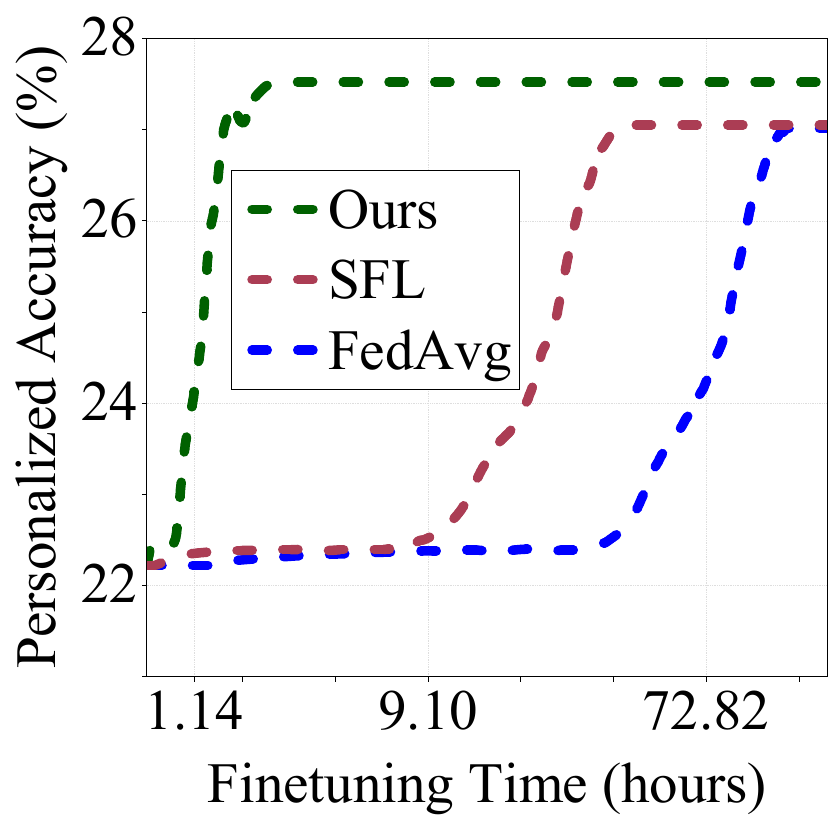}
      \caption{High School Psychology}
      \label{fig:gptmediumpsychology}
    \end{subfigure}
    \hfill
    \begin{subfigure}[t]{0.47\linewidth}
      \centering
      \includegraphics[width=\linewidth]{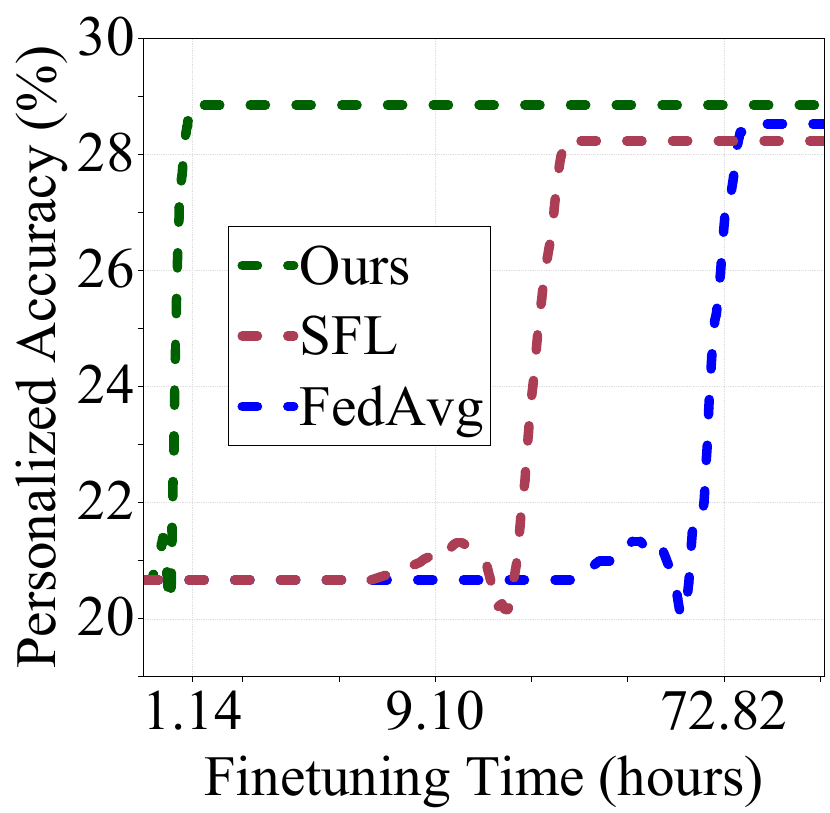}
      \caption{Professional Law}
      \label{fig:gptmediumlaw}
    \end{subfigure}

    \caption{Client personalized performance on different tasks.}

    \label{fig:personalized}
\vspace{-4mm}
\end{figure}

\
\newline
\indent \textbf{Personalized accuracy \& fine-tuning time:} First, we focus on personalized accuracy and the fine-tuning time refers to the system wall-clock time. Compared with SFL and FedAvg, we make the following observations:
\begin{itemize}
\item As shown in Table~\ref{tab:overall}, FlexP-SFT outperforms both FedAvg and SFL in terms of final average personalized accuracy and overall fine-tuning time.

\item Figure~\ref{fig:personalized} further illustrates the personalized performance on a representative client for the GPT-Medium@ MMLU task. For example, in the High School Psychology task, FlexP-SFT achieved a final accuracy of 27.49\%, compared to 26.90\% for FedAvg and 27.04\% for SFL.federated settings.
\end{itemize}

\textbf{Memory Consumed \& Communication Throughput:} Figure~\ref{fig:localmemory} \&~\ref{fig:commthroughput} further illustrates the memory consumption and total communication throughput with GPT2-Small@ MMLU.
\begin{itemize}
\item As shown in Fig.~\ref{fig:localmemory}, although the memory usage of our method increases with \( Q \) due to the addition of layers, it remains substantially lower than the other two methods. SFL requires additional memory space to package local model parameters for transmission to the server. FedAvg, being a full-model approach, consumes the largest amount of memory. 
\item In Fig.~\ref{fig:commthroughput}, we compare the total communication throughput with FedAvg and SFL. Since our method transmits only activations and gradients, the single package size is approximately 3MB, significantly smaller than the model parameters (GBs) required by the other methods. Moreover, the communication throughput of our method remains constant regardless of the local model size \( Q \). This results in our method consuming far less total communication bandwidth compared to FedAvg and SFL.
\end{itemize}

\subsubsection{Comparison with PFL methods}
\
\newline
\indent \textbf{Personalized accuracy \& fine-tuning time:}
We compare our method with state-of-the-art PFL approaches. The results are summarized in Table~\ref{tab:personalized_performance}. 
\begin{itemize}
\item FlexP-SFT drastically reduces fine-tuning time with comparable or superior personalized accuracy.
\item The efficiency gain does not come at the expense of accuracy: FlexP-SFT delivers the highest accuracy on BERT-Base (28.81\%), and remains competitive with or better than all baselines across other models.
\end{itemize}

\subsubsection{Comparison with federated on-device fine-tuning of
LLM method}
\
\newline
\indent The methods in this baseline group differ significantly in several aspects, including the number of trainable parameters (e.g., only a few parameters are updated in LoRA-based methods), reliance on server-side computation (e.g., adapter-based methods delegate most of the computation to the server), and gradient update strategies (e.g., forward-propagation-based methods acquire gradients very slowly). Thus, we argue that comparing fine-tuning time with them is not fair. In this section, we focus on comparing final personalized accuracy, memory consumption, and communication throughput. As shown in the Tab.~\ref{tab:3rdbaseline}, we make the following observations:
\
\newline
\indent \textbf{Personalized Accuracy:}
\begin{itemize}
\item FlexP-SFT achieves a higher average final personalized accuracy than other baselines, primarily due to its personalized model design.
\end{itemize}

\begin{table}[]
\centering
\caption{Impact of $Q$ on final personalized accuracy after model convergence.}
\label{tab:insight2}

\resizebox{\linewidth}{!}{%
\begin{tabular}{clcccc}
\toprule
\multicolumn{3}{c}{\textbf{Settings}} & \multicolumn{3}{c}{\textbf{Final Accuracy (\%)}} \\ 
\cmidrule(lr){1-3} \cmidrule(lr){4-6}
\multicolumn{1}{c}{Model} & \#Para & Dataset & $Q=0.1$ & $Q=0.2$ & $Q=0.5$ \\ 
\midrule
BERT-Large     & 340M  & \multirow{4}{*}{MMLU}     & 30.02 & 31.00 & 31.94 \\
BERT-Base      & 110M  &                          & 28.85 & 28.33 & 31.69 \\
GPT2-Medium    & 355M  &                          & 25.68 & 27.49 & 28.67 \\
GPT2-Small     & 124M  &                          & 22.05 & 25.15 & 28.67 \\
\midrule
BERT-Large     & 340M  & \multirow{4}{*}{MMLU-Pro} & 18.45 & 20.89 & 22.95 \\
BERT-Base      & 110M  &                          & 18.94 & 18.89 & 20.14 \\
GPT2-Medium    & 355M  &                          & 13.23 & 15.20 & 17.60 \\
GPT2-Small     & 124M  &                          & 10.14 & 10.93 & 11.73 \\
\bottomrule
\end{tabular}%
}
\end{table}

\begin{table}[]
\centering
\caption{Hyper-parameters for resource-aware split-ratio selection.}
\label{tab:q_hyperparameters}
\resizebox{\linewidth}{!}{%
\begin{tabular}{lll}
\toprule
Parameter & Symbol & Value \\
\midrule
Warm-up steps & $K_p$ & 50 \\
Validation split & $|\mathcal{D}_n^{\mathrm{val}}|$ & 10\% of local data \\
Numerical constant & $\epsilon$ & $10^{-8}$ \\
Alignment weight & $\lambda$ & 0.1 \\
Memory/time weights & $\omega_M,\omega_T$ & 0.5, 0.5 \\
Default trade-off & $\alpha$ & 0.7 \\
Trade-off sweep & $\alpha$ & $\{0.1,0.3,0.5,0.6,0.7,0.8,0.9\}$ \\
\bottomrule
\end{tabular}

}
\vspace{-2mm}
\end{table}

\textbf{Memory Consumed \& Communication Throughput:}
\begin{itemize}
    \item FlexP-SFT consumes less local memory compared to other baselines, as it does not require storing or loading the backbone model on the client side.
    \item FlexP-SFT consumes less communication throughput on the device, as it avoids transmitting large model parameters (unlike PocketLLM) and only needs to transmit activations—approximately 5× activations' size, which is still lower than methods like MobiLLM that transmit activations after every layer.
\end{itemize}

\begin{table*}[]
\centering
\caption{Split-ratio selection on the heterogeneous six-client testbed. Optimized Q automatically selects client-specific ratios according to the user-specified accuracy--resource preference ($\alpha=0.7$). In the selected-\(Q_{1:6}\) column, C1--C2 are Xavier devices, C3--C4 are Nano devices, C5 is the Raspberry Pi 5, and C6 is the Raspberry Pi 4B; ``-'' denotes an infeasible static request.}
\label{tab:q_selection}
\resizebox{\textwidth}{!}{%
\begin{tabular}{llllllll}
\toprule
Strategy & Alpha & Selected $Q_{1:6}$ & Accuracy (\%) & Fine-tuning time (h) & Avg. memory (GB) & Avg. energy (J/step) & Dropout rate (\%) \\
\midrule
Static Q=0.1 & - & 0.1, 0.1, 0.1, 0.1, 0.1, 0.1 & 25.68 & 1.53 & 1.23 & 32.50  & 0.00 \\
Static Q=0.3 & - & 0.3, 0.3, 0.3, 0.3, 0.3, 0.3 & 27.88 & 3.28 & 2.43 & 79.44  & 0.00 \\
Static Q=0.5 & - & 0.5, 0.5, -, -, 0.5, - & 26.68 & 4.13 & 3.57 & 125.79 &  50.00 \\
Memory-constrained Max-Q & - & 0.5, 0.5, 0.3, 0.3, 0.5, 0.3 & 28.28 & 3.81 & 3.02 & 97.04  & 0.00 \\
Optimized Q & 0.70 & 0.5, 0.5, 0.2, 0.2, 0.5, 0.2 & 28.08 & 3.50 & 2.78 & 90.39  & 0.00 \\
\bottomrule
\end{tabular}

}
\vspace{-2mm}
\end{table*}

\begin{figure}[t]
  \centering
  \begin{subfigure}[t]{0.48\linewidth}
    \centering
    \includegraphics[width=\linewidth]{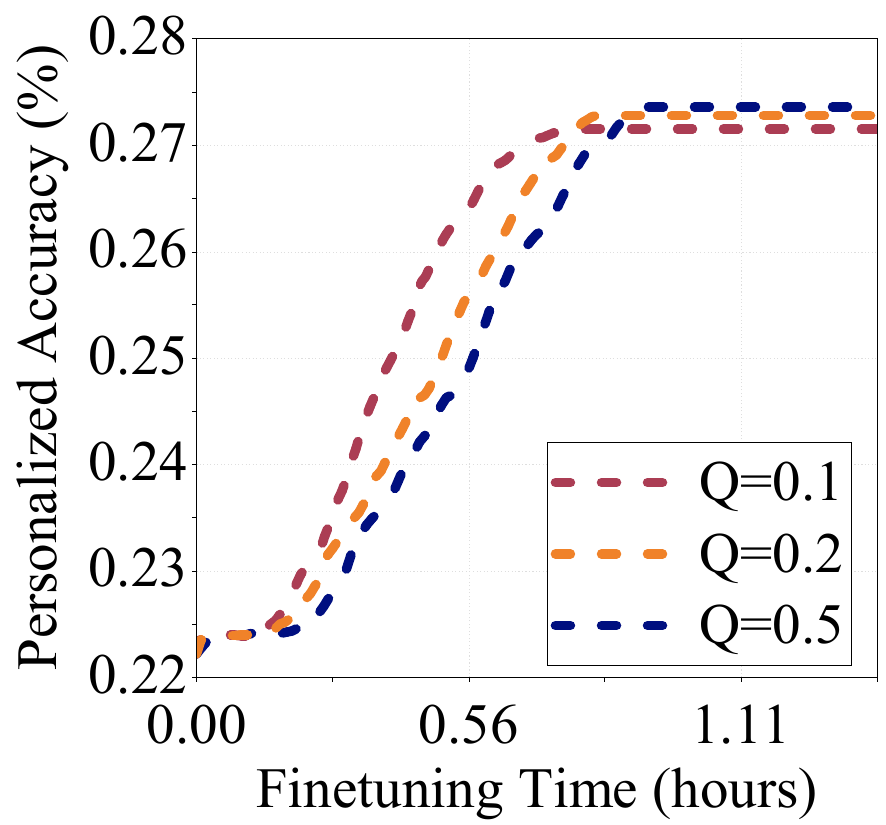}
    \caption{Analysis \( Q \)}
    \label{fig:accq}
  \end{subfigure}
  \hfill
  \begin{subfigure}[t]{0.49\linewidth}
    \centering
    \includegraphics[width=\linewidth]{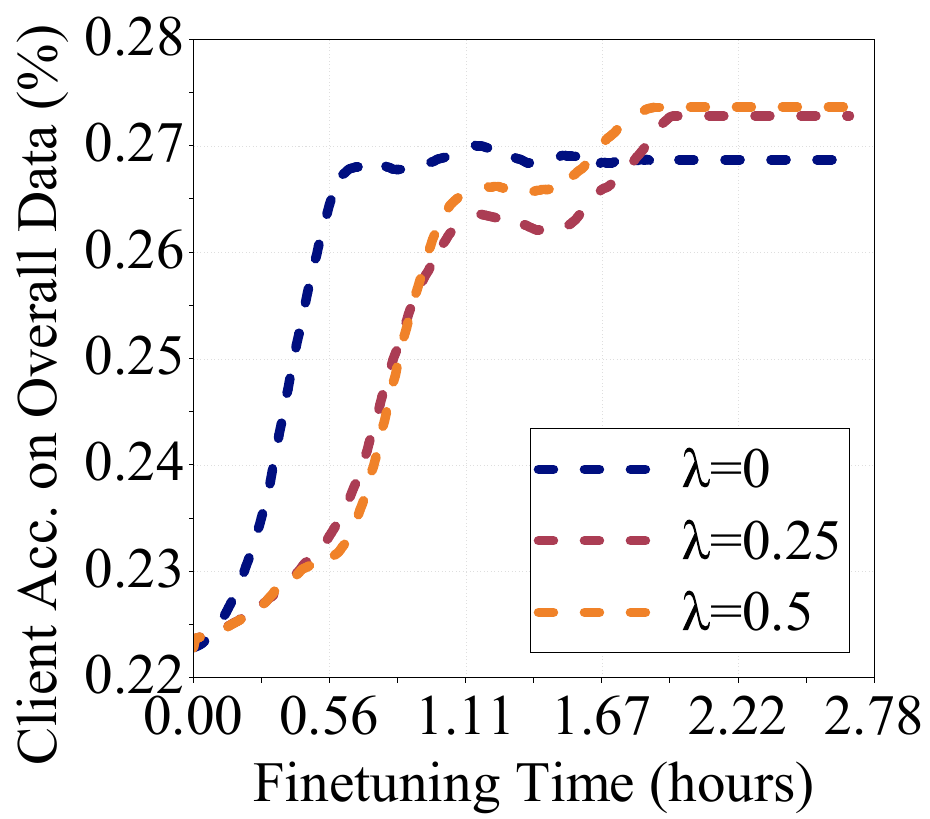}
    \caption{Analysis $\lambda$}
    \label{fig:lambda}
  \end{subfigure}
  \caption{Ablation study of \( Q \) and $\lambda$.}
  \label{fig:analysis_q_lambda}
\vspace{-4mm}
\end{figure}

\begin{figure}[t]
\centering
\includegraphics[width=0.8\linewidth]{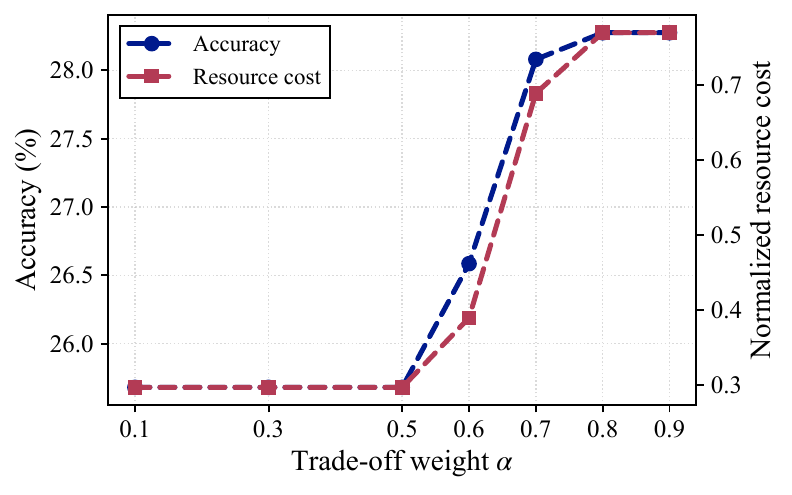}
\caption{Sensitivity to the trade-off weight $\alpha$. Increasing $\alpha$ prioritizes the warm-up personalization surrogate and raises both final accuracy and normalized client-side resource cost.}
\label{fig:alpha_tradeoff}
\vspace{-3mm}
\end{figure}

\begin{table}[]
\centering
\caption{Impact of Client Dropout on Final Personalized Accuracy and Fine-tuning Time.}
\label{tab:dropout}
\resizebox{0.85\linewidth}{!}{%
\begin{tabular}{ccc}
\toprule
\textbf{Dropout Rate (\%)} & \textbf{Accuracy (\%)} & \textbf{Fine-tuning Time (h)} \\
\midrule
0   & 26.20 & 1.63 \\
10  & 26.11 & 2.09 \\
50  & 25.65 & 2.17 \\
\bottomrule
\end{tabular}%
}
\end{table}

\subsection{Ablation Study}

\subsubsection{Flexible Layers}

Fig.~\ref{fig:accq} shows the relationship between model accuracy and wall-clock time on the same client using different values of $Q$, with all other factors held constant, including the client device (Xavier), local dataset, and task (Professional Law Task from MMLU). To further investigate the impact of $Q$ on final accuracy, we conducted additional experiments summarized in Tab.~\ref{tab:insight2}. 

\textbf{Effect of Q on personalization.}
\begin{itemize}
\item The results consistently show that higher values of $Q$ lead to better final personalized accuracy. \item Increasing $Q$ offloads more computation to the client, resulting in slower training but improved personalization. 
Smaller $Q$ values shift computation to the server, achieving faster training at the cost of reduced personalized performance. 
\end{itemize}

\subsubsection{Resource-Aware Q Selection}

We next evaluate whether Problem~(P1) and Algorithm~\ref{alg:q_selection} produce a better split choice than fixed or single-constraint rules. For each of the six clients, the profiler enumerates $Q_n\in\{0.1,0.2,0.3,0.5\}$, measures $M_n(Q_n)$ and $T_n(Q_n)$, and computes $\widehat{A}_n(Q_n,\lambda)$ from the relative validation-loss reduction after the 50-step warm-up described in Sec.~IV-A.

We compare five strategies. Static $Q=0.1$, Static $Q=0.3$, and Static $Q=0.5$ force all clients to request the same split. Memory-constrained Max-Q selects, for each client, the largest $Q_n$ satisfying only its memory budget and ignores latency and $\widehat{A}_n$. Optimized Q instead performs demand-aware automatic selection. 
Table~\ref{tab:q_selection} exposes the limitations of fixed choices and shows how the automatic selector balances competing demands. Static $Q=0.1$ gives the lowest time and energy, but its accuracy is only 25.68\% and its system cost is the highest. Static $Q=0.5$ retains more personalized layers, but half of the clients cannot satisfy their local budgets; its apparently low system cost therefore comes at a 50\% dropout rate. Memory-constrained Max-Q avoids dropout and reaches 28.275\% accuracy, but it always consumes the largest memory-feasible local model without considering whether the extra local cost is justified by its personalization gain. Under the balanced user preference $\alpha=0.7$, Optimized Q automatically assigns $Q_n=0.5$ to the Xavier pair (C1--C2) and the Raspberry Pi 5 (C5), while assigning $Q_n=0.2$ to the Nano pair (C3--C4) and the Raspberry Pi 4B (C6). Optimized Q selects an appropriate feasible split for each client under the user's requested accuracy--resource trade-off.

Fig.~\ref{fig:alpha_tradeoff} further verifies the role of $\alpha$. To make resource usage comparable across clients, the right y-axis reports the average normalized resource cost $\overline{C}_{\mathrm{norm}}(\mathbf{Q})
=\frac{1}{N}\sum_{n=1}^{N}\frac{C_n(Q_n)}{C_{\max}},$
where $C_n(Q_n)$ is the weighted memory--latency cost defined in Sec.~III-C and $C_{\max}$ is its maximum admissible value. Because feasible configurations satisfy $M_n(Q_n)/\overline{M}_n\leq1$ and $T_n(Q_n)/\overline{T}_n\leq1$, and because $\omega_M+\omega_T=1$, we have $C_{\max}=\omega_M+\omega_T=1$. Thus, a value closer to 1 indicates that the selected split consumes a larger fraction of the clients' available resource budgets. When $\alpha\leq0.5$, the resource term dominates and all clients select $Q_n=0.1$, yielding 25.68\% accuracy and an average normalized resource cost of 0.297. At $\alpha=0.6$, capable clients move to $Q_n=0.2$, increasing accuracy to 26.585\%. The default $\alpha=0.7$ yields the heterogeneous balanced solution above and reaches 28.08\% accuracy at a normalized resource cost of 0.689. When $\alpha\geq0.8$, each client selects its largest feasible split, increasing accuracy to 28.275\% but also increasing the normalized resource cost to 0.769. The stepwise curve is expected because $Q_n$ is a discrete layer-level decision. These results directly validate that $\alpha$ controls the intended accuracy--resource trade-off in Problem~(P1).

\begin{table}[]
\centering
\caption{Individual Model Accuracy on Local and Other Clients’ Data Across Different Personalization Levels. (GPT2-Medium@MMLU)}
\label{tab:person}

\resizebox{\linewidth}{!}{%
\begin{tabular}{cccccc}
\toprule
\textbf{Settings} & \multicolumn{5}{c}{\textbf{Client 3's Model Test Accuracy (\%) on Data from:}} \\
\cmidrule(lr){2-6}
$\lambda$ & Client 1 & Client 2 & \textbf{Client 3 (Personalized)} & Client 4 & Client 5 \\
\midrule
0     & 22.95 & 26.12 & \textbf{28.33} & 27.38 & 24.27 \\
0.25  & 20.77 & 26.51 & \textbf{29.02} & 28.14 & 25.24 \\
0.5   & 21.31 & 26.90 & \textbf{29.00} & 27.38 & 25.73 \\
\midrule
\textbf{Settings} & \multicolumn{5}{c}{\textbf{Client 3's Data Test Accuracy (\%) on Models from:}} \\
\cmidrule(lr){2-6}
$\lambda$ & Client 1 & Client 2 & \textbf{Client 3 (Personalized)} & Client 4 & Client 5 \\
\midrule
0.25  & 23.79 & 26.33 & \textbf{29.02} & 21.83 & 19.39 \\
\bottomrule
\end{tabular}%
}
\vspace{-4mm}
\end{table}

\subsubsection{Alignment Strategy}
\
\newline
\indent Figure~\ref{fig:lambda} illustrates the performance of the client model on the global dataset for different values of \(\lambda\) (0, 0.25, and 0.5). 
\begin{itemize}
\item When \(\lambda = 0\), our alignment strategy is not used, resulting in faster convergence but lower final accuracy. 
\item As \(\lambda\) increases, the final accuracy improves. This is because a larger \(\lambda\) means the client model becomes more similar to the global model, enabling the client to benefit from knowledge learned on other devices. This knowledge sharing enhances the model's accuracy on tasks across different clients, leading to better overall performance. 
\end{itemize}

\subsection{Robustness}
\subsubsection{Client Dropout}
\
\newline
\indent We train GPT2-Medium on the MMLU dataset with FlexP-SFT under different dropout rates: 0\% dropout, where all clients participate in every round; 10\% dropout, where each client has a 10\% chance of dropping out in any round; and 50\% dropout, where each client has a 50\% chance of dropping out in any round. The results in Tab.\ref{tab:dropout} show that:
\begin{itemize}
\item As the dropout rate increases, the final average personalized accuracy experiences a slight decrease. 
\item The overall fine-tuning time increases from 1.63 hours to 2.17 hours as dropout becomes more frequent. This indicates that while higher client dropout slightly impacts the personalization performance, it also leads to longer fine-tuning durations due to the additional training steps needed to compensate for the missing updates.
\end{itemize}

\subsubsection{How Personalization Adapts to Individual Client}
\
\newline
\indent We choose client 3 as an example. As shown in Tab.~\ref{tab:person}, we make the following observations: 
\begin{itemize}
\item Increasing the personalization coefficient $\lambda$ slightly improves local performance.

\item Client 3's model evaluated with data on others: Client 3's model consistently achieved the highest accuracy on its own data. 

\item Each client’s final model evaluated on Client 3’s data:
Client 3’s own model attains the highest accuracy. 
\end{itemize}
These effectively validate the personalization ability of FlexP-SFT.

\section{Conclusion}

In this paper, we introduced FlexP-SFT, a novel aggregation-free framework designed for personalized split federated fine-tuning of LLMs on resource-constrained edge devices. By fundamentally eliminating the dependency on client-side parameter aggregation, FlexP-SFT enhances personalized performance, reduces communication overhead, and resolves the straggler promblem. In addition, we proposed a layer-flexible alignment strategy for balancing personalization and generalization to address the challenges of heterogeneity and local overfitting. We further formulated the client-side split ratio as a resource-aware discrete multi-variable optimization problem. While the scale and structural complexity of LLMs pose challenges for rigorous theoretical convergence analysis, FlexP-SFT achieves strong empirical results and offers a practical solution for on-device fine-tuning. We believe these contributions promote the real-world deployment of personalized AI on edge devices.


\bibliography{old/ref}
\bibliographystyle{IEEEtran}


\end{document}